%% file: main.tex
\documentclass[11pt]{article}
\usepackage[top=0.8in, bottom=0.78in, left=0.87in, right=0.87in]{geometry}
\usepackage{setspace}
\usepackage[T1]{fontenc}
\usepackage{times}
\usepackage{booktabs}
\usepackage{rotating}
\usepackage{longtable}
\usepackage{graphicx}
\usepackage{tikz}
\usepackage[section]{placeins} 
\usepackage[large, bf]{caption}
\usepackage[FIGTOPCAP]{subfigure}
\usepackage{palatino}
\usepackage{textcomp}
\usepackage{longtable}
\usepackage{nicefrac}
\usepackage{adjustbox}    
\usepackage[hyphens]{url}
\usepackage{natbib}
\usepackage{float}
\bibpunct{(}{)}{;}{a}{,}{,}

\usepackage{hyperref}
\hypersetup{
    colorlinks=true,
    linkcolor=blue, 
    citecolor=blue, 
    urlcolor=blue, 
}

\usepackage{amsmath, amsfonts, amssymb, amsthm}
\usepackage{mathpazo}

\usepackage{mdframed}
\usepackage{enumitem}

\def\urltilda{\kern -.15em\lower .7ex\hbox{\~{}}\kern .04em}

\theoremstyle{plain}  

\theoremstyle{definition}  

\theoremstyle{remark}  

\usepackage{titlesec}

\titlespacing*{\section}{0pt}{1.5ex plus 1ex minus .2ex}{0.8ex plus .2ex}
\titlespacing*{\subsection}{0pt}{1.2ex plus 1ex minus .2ex}{0.8ex plus .2ex}

\begin{document}

\title{The Value of Non-Traditional Credentials in the Labor Market\thanks{Athey: Stanford University; Palikot: Northeastern University. We thank Eric Karsten, Yixi Jiang, and Anna Zhao from Coursera for collaborating on this project. We thank Keshav Agrawal and Elena Pittarokoili for excellent research assistance. The Golub Capital Social Impact Lab at Stanford Graduate School of Business provided funding for this research. This research has been subject to review and approval by Research Compliance Office at Stanford University, protocol number IRB-59983 and registered at AEA RCT registry AEARCTR-0009438
.}}

\author{Susan Athey \& Emil Palikot}
\date{\today}
\maketitle

\begin{abstract}
\noindent
Workers without formal credentials experience substantially lower employment rates than their credentialed counterparts, but the extent to which information frictions contribute to these disparities remains unclear. We conducted a randomized experiment with over 800,000 online certificate earners from developing countries who lack college degrees, encouraging credential sharing on LinkedIn through reduced friction and reminders. We study credential visibility for the full sample and track employment outcomes for 40,000 learners. The intervention increased new employment by 6\% (1.0 percentage point), with larger effects of 9\% (1.2 percentage points) for jobs related to certificates. Treatment effects concentrate among learners with weak baseline employability: those in the bottom tercile experience employment gains of 12\% while those in the top tercile see negligible effects. Those most responsive to the intervention benefit most from sharing, with a local average treatment effect of 11 percentage points for compliers. Counterfactual resume scoring shows that credentials improve perceived candidate quality by 8.5 points on average (on a scale from 1 to 100), with effects of 15-20 points for weak resumes but near-zero for strong resumes. Our findings demonstrate that information frictions substantially constrain employment for workers without traditional credentials, and that minimal-cost interventions targeting credential visibility can generate employment gains comparable to intensive training programs.
\end{abstract}

\newpage

\input{1.intro}
\input{2.empirical_setting}
\input{4.main_result}

\input{5.resume_scoring}
\input{6.conclusion}
\newpage

\bibliographystyle{apalike}
\bibliography{refs}

\input{appendix}

\end{document}

%% file: 1.intro.tex
\section{Introduction}

Workers lacking formal educational credentials experience substantially lower employment rates and earnings than their credentialed counterparts \citep{card1999causal, goldin2018race}. A key question in designing policies to improve these workers' labor market outcomes is whether their poor outcomes result from skill deficiencies or from information frictions that prevent employers from observing their existing capabilities. If it is the former, then these workers may need to engage in intensive training programs to acquire marketable skills. If it is the latter, then interventions that simply make workers' skills visible to potential employers, such as encouraging credential sharing on professional networking platforms, may have substantial benefits at minimal cost.

In today's educational landscape, non-traditional credentials acquired through online courses have become increasingly popular. In 2024, Coursera, one of the largest providers of Massive Open Online Courses (MOOCs), reported 162 million registered learners \citep{coursera2024}. Many learners gain these credentials with the intention of signaling specific skills to potential employers \citep{laryea2021ambiguous}. Despite their prevalence, there remains a significant gap in high-quality evidence demonstrating their actual value in the labor market, particularly for learners who lack access to traditional credentials. This raises important questions: To what extent do these non-traditional credentials help learners secure new employment? Moreover, who benefits the most from them? And through what mechanism do credentials affect hiring outcomes?

We conducted a randomized experiment with over 800,000 learners on Coursera, a prominent MOOC provider, encouraging them to add newly earned credentials to their LinkedIn profiles. We focus on learners who recently completed career-oriented certificate programs and who either lack college degrees or come from developing countries, populations that often lack access to traditional credentials or internationally renowned educational institutions \citep{hansen2015democratizing, moura2017moocs}. A randomly selected subset received access to the \textit{Credential Feature}, which provided a simplified process for showcasing Coursera credentials on LinkedIn profiles, along with targeted notifications encouraging this action. The control group did not receive access to the feature. For the full experimental sample of 800,000 learners, we observe whether their credentials attract views on LinkedIn, providing a measure of credential visibility to potential employers. Our primary analysis focuses on approximately 40,000 learners who provided LinkedIn profile URLs in their Coursera accounts before randomization, allowing us to track not only credential visibility but also whether they reported new employment after the intervention.

The intervention generated substantial employment gains. Learners in the treatment group are 5.9\% (S.E. 2.6\%) more likely to find new employment within a year, representing a 1.0 percentage point increase from the baseline of 17.3\%. The effect is larger for jobs directly related to the certificate: a 9.4\% (S.E. 3.1\%) higher likelihood, an increase of 1.2 percentage points from the baseline of 12.7\%. These results remain robust when excluding jobs reported with starting dates within the first four months after randomization, addressing concerns that treatment might simply accelerate profile updates rather than generate new employment. The effect appears primarily mediated by credential presence: the treatment group is 17.3\% (S.E. 2.4\%) more likely to share their credentials on LinkedIn. Critically, treatment effects concentrate among learners with low and moderate baseline employability, demonstrating that credentials provide the greatest value precisely where traditional signals are weakest.

We can also interpret the experiment as an encouragement design, with random assignment of the \textit{Credential Feature} serving as an instrument for credential sharing. First, we find substantial heterogeneity in who responds to the encouragement. Using Rank Average Treatment Effects (RATE), we show that targeting the top 20\% of learners leads to 50\% higher estimates of the average effect on credential sharing than targeting the 40\% of learners with the highest predicted treatment effect. Learners most responsive to the intervention are those with lower skill scores, from developing countries, in IT domains, and without prior technical employment. Second, we estimate Local Average Treatment Effects (LATE) separately for learners from high and low RATE groups. Learners most responsive to our treatment have an 11 percentage point (S.E. 2.8 percentage points) local average treatment effect, while the low RATE group's local average treatment effect is low and statistically insignificant. This alignment between treatment responsiveness and treatment benefits has implications for efficient targeting: platforms can identify high-value populations using observable characteristics and focus their efforts on encouraging these learners to use their credentials.

To understand the mechanism underlying these employment effects, we conduct a resume scoring analysis using an LLM-as-judge framework. We simulate recruiter evaluation by scoring LinkedIn profiles with and without credentials using Resume-Matcher \citep{resume_matcher}, an open-source tool that prompts a large language model to act as a recruiter and assess candidate-job fit on a 0-100 scale. Credentials improve resume scores by an average of 8.5 points. Critically, this effect is highly heterogeneous: learners with the weakest baseline resumes (scores below 20) experience improvements of 15 to 20 points, while those with strong existing signals (scores above 70) see minimal or even negative effects. We validate that resume scores predict actual employment outcomes in our control group, and than find that the mean improvement of 8.5 points translates to a 5.1 percentage point higher chance of finding a new job. The inverse relationship between baseline resume quality and credential value provides direct evidence that credentials function as salience-enhancing signals, making candidates more visible during screening precisely when they lack traditional credentials.

A secondary analysis examines surrogate outcomes for the entire experimental population of over 800,000 learners. Although we do not observe LinkedIn profiles for most learners, we measure whether credentials received clicks from other LinkedIn users, potentially indicating employer interest. Clicks strongly correlate with new job reporting in the subsample with employment outcomes. Adjusted for learner characteristics, we estimate that the \textit{Credential Feature} increases the probability of receiving credential views by 2\% to 4\% from the baseline probability of 13\%, suggesting that the intervention's impact extended beyond the subsample where we observed employment outcomes. The substantially lower treatment effects and baseline rates in the full sample compared to the LinkedIn-matched subsample indicate that learners who actively maintain LinkedIn profiles are both more responsive to credential-sharing interventions and benefit more from them. This pattern also suggests that a meaningful portion of MOOC learners may not actively use professional networking platforms, limiting their ability to benefit from credential visibility interventions regardless of treatment assignment.

Our findings directly address the question posed at the outset: for workers without traditional credentials, poor labor market outcomes stem also from information frictions rather than skill deficiencies alone. A light-touch platform intervention with minimal implementation costs generated employment gains comparable to those documented for intensive training programs. Interventions that make existing capabilities visible, by increasing the salience of credentials, reducing sharing friction, and leveraging platform credibility, can generate meaningful employment gains at substantially lower cost than skill-building programs. For educational platforms and policymakers, this suggests an alternative approach to reducing employment disparities: rather than requiring workers to acquire new skills through costly programs, platforms can facilitate workers' ability to effectively signal their existing capabilities at scale. The concentration of treatment effects among disadvantaged learners indicates that facilitating credential sharing may reduce rather than exacerbate employment inequality.

\section{Literature Review}

This paper contributes to the large literature in economics on labor market signaling and educational credentials \citep{spence1978job, tyler2000estimating, hussey2012human}. Recent work has examined how workers signal capability through non-traditional credentials when formal educational signals are unavailable or insufficient. \citet{pallais2014inefficient} shows that on oDesk (now Upwork), credentials and work history serve as valuable signals that reduce information frictions in hiring, with particularly large effects for workers entering the platform. \citet{kassi2019digital} find that digital skill certificates help new workers enter online labor markets by providing verifiable signals of competence. Several field experiments demonstrate that interventions facilitating skill signaling can improve employment outcomes in developing countries. \citet{abebe2020matching} evaluate a job application workshop in Ethiopia that provides certificates in various skills, finding improvements in employment and significant earnings increases. \citet{carranza2020job} show that certificates improved job search outcomes and increased callbacks from firms in South Africa. \citet{bassi2022screening} demonstrate that soft skills certificates in Uganda increased employability and earnings, with effects driven by reduced screening costs for employers. \citet{piopiunik2020skills} provide experimental evidence from Germany that skill signals significantly affect job interview invitations, with effects varying by signal type and labor market context. \citet{athey2022effective} document large impacts of a program helping women develop technical portfolios, showing that facilitating skill visibility can improve employment outcomes for disadvantaged groups. 

Our study contributes to this literature by examining credential-sharing interventions at scale on mainstream professional networking platforms, focusing on heterogeneity in treatment effects across baseline employability levels. While prior work establishes that non-traditional credentials can improve outcomes, we show that their value is highly heterogeneous and concentrates among workers with the weakest alternative signals. Moreover, we demonstrate that those most responsive to credential-sharing nudges are precisely those who benefit most from sharing, providing actionable insights for platform design and targeting.

Survey evidence reveals mixed employer perceptions of MOOC credentials. \citet{rosendale2016valuing} surveyed 202 employers, finding a general preference for traditional degrees over MOOC credentials. \citet{radford2014employer} surveyed 103 human resources professionals and found that while MOOCs were viewed favorably on resumes, they were perceived as less likely to demonstrate specific skills than traditional credentials. \citet{kizilcec2019online} document that respondents view online degree programs as less legitimate and respected than conventional degrees.

These survey-based studies provide important context about stated preferences, but leave open whether employer perceptions translate into actual hiring decisions. Our experimental evidence addresses this gap by measuring real employment outcomes, suggesting that despite stated skepticism, MOOC credentials do attract employer interest and generate employment gains, particularly for workers lacking traditional credentials.

Randomized audit studies, sending fictitious resumes to real job postings, provide causal evidence on how employers respond to online credentials. \citet{deming2016value} and \citet{lennon2021online} show that online degrees generate fewer callbacks than traditional degrees from comparable institutions, suggesting employer discrimination against online credentials. \citet{rivas2020moocs} use a recruited experiment where Mechanical Turk workers select among hypothetical profiles, demonstrating that MOOC credentials increase selection probability compared to having no credentials at all.

Observational and quasi-experimental studies examine outcomes for MOOC completers. \citet{hadavand2018can} compare completers to non-completers of a data science specialization, estimating salary increases and higher job mobility for completers. \citet{zhenghao2015whos} survey MOOC participants after course completion and find that a majority report career benefits. However, these studies face selection concerns: completers may differ from non-completers in unobserved dimensions such as motivation or baseline ability.

Two recent randomized experiments attempt to overcome these selection issues by randomizing incentives to enroll in MOOCs. \citet{novella2024online} study a Coursera program in Costa Rica, while \citet{majerowicz2023massive} examine MOOC access in Colombia. Both experiments provide financial incentives for enrollment. However, low course completion rates (typically below 10\%) substantially reduce statistical power, rendering estimates of average treatment effects on employment outcomes inconclusive.

Our work advances this literature in several ways. First, we conduct a large-scale randomized experiment focusing on course completers, achieving sufficient statistical power to detect employment effects and analyze heterogeneity. Second, rather than randomizing skill acquisition (enrollment in courses), we randomize credential visibility, isolating the signaling channel from human capital formation. Third, we track actual employment outcomes on LinkedIn rather than relying on survey self-reports or audit study callbacks. Fourth, we demonstrate that MOOC credential value is highly heterogeneous, reconciling seemingly conflicting findings: credentials provide minimal value for workers with strong traditional signals but generate substantial employment gains for those lacking credentials.

Our mechanism analysis employs an emerging methodology in which large language models serve as evaluators of complex artifacts—an approach known as "LLM-as-a-judge" \citep{zheng2023judging}. This framework has gained traction as a scalable alternative to human evaluation for tasks where obtaining expert assessments is costly or time-consuming. \citet{zheng2023judging, bommasani2021foundation} demonstrate that LLM-based evaluators can achieve high agreement rates with human judges across diverse evaluation tasks, with particularly strong performance on structured assessment criteria.

Recent work extends this approach to economic contexts. \citet{horton2023large} shows that large language models can replicate complex human judgments in economic settings, including evaluating worker quality and task performance in online labor markets, with performance approaching that of expert raters. \citet{manning2025general} demonstrate that AI agents can predict human behavior in novel settings by combining theory-grounded instructions with existing empirical data, even outperforming traditional behavioral models when predicting responses to new scenarios.

The key advantage of LLM-as-judge frameworks for our research question lies in their ability to provide consistent, reproducible evaluations at scale while simulating realistic screening processes. Unlike traditional resume parsing or keyword matching, modern LLMs can assess semantic fit between candidate qualifications and job requirements \citep{horton2023large, vafa2022career, manning2025general}.

We contribute to this methodological literature by demonstrating how LLM-based resume scoring can provide evidence for economic mechanisms in field experiments. Rather than treating LLM evaluations as ground truth for hiring outcomes, we use them as a measure of perceived candidate quality, a proxy for how credentials affect initial screening visibility. Our validation analysis demonstrates strong correlations between LLM-generated scores and actual employment outcomes in our experimental sample, supporting the validity of this approach for measuring the salience mechanism through which credentials affect employment.

%% file: 2.empirical_setting.tex
\section{Empirical Setting and Experimental Design}
\subsection{The Coursera Platform and Credentialing Ecosystem}
Coursera, one of the largest online platforms hosting MOOCs, is characterized by its extensive course offerings and partnerships with global universities and organizations. In 2024, the platform reported 162 million registered learners, adding more than 26 million new learners during the year \citep{coursera2024}. The platform's business model combines free access with paid credentials: most courses can be audited for free, but obtaining a certificate typically involves a fee ranging from \$29 to \$99 for individual courses. Specializations and professional certificates, which consist of a series of related courses, usually cost between \$39 and \$79 per month, with the total expense depending on the time taken to complete the series.\footnote{Coursera also offers online degrees with significantly higher costs, but individuals graduating with online degrees are not part of this study.} The affordability and flexibility of Coursera's offerings are central to its appeal, particularly for learners from economically disadvantaged regions or marginalized groups \citep{kizilcec2017closing, chirikov2020online}.

Many courses offered by Coursera allow learners to obtain completion certificates. In addition to paying for them, obtaining certificates typically requires completing coursework and passing assessments. These certificates are often valued for their focus on practical skills relevant to career advancement, and observational data studies and recruited experiments suggest that credentials obtained through such courses can positively impact career progression \citep{hadavand2018can, rivas2020moocs, castano2021open}. Many Coursera courses are thus career-oriented, and some of the most popular domains include \emph{Information Technology}, \emph{Computer Science}, \emph{Data Science}, and \emph{Business}.

While Coursera issues credentials upon course completion, learners must take additional steps to display these credentials on professional networking platforms like LinkedIn, where they can be discovered by potential employers. This process historically required learners to copy a credential URL from their Coursera certificate page and paste it into their LinkedIn profile, which is a multi-step process requiring several clicks and navigation across platforms. 

In the sample of learners who reported LinkedIn accounts upon registration for a course on Coursera, only round 18 percent added their credential to the LinkedIn profiles, suggesting substantial friction or uncertainty about the value of doing so. Prior research documents comparable patterns of undersharing of professional credentials \citep{laryea2021ambiguous}.

Several mechanisms might contribute to this low baseline. First, adding credentials to LinkedIn involves friction costs from a multi-step process requiring navigation across platforms, copying of URLs, and form completion. Such frictions can prevent value-creating activities from occurring when participants face imperfect information about returns \citep{williamson1985economic}. Empirical evidence demonstrates that even modest increases in process complexity can substantially reduce completion rates. \citet{madrian2001power} and \citet{carroll2009optimal} show that changing 401(k) enrollment from an opt-in process (requiring employees to actively complete enrollment forms) to automatic enrollment substantially increases participation, highlighting the power of reducing even small activation costs. 
Second, the existing platform design does not give high salience to credential sharing. Research on salience demonstrates that decision-makers focus attention on stimuli that are surprising, prominent, or presented at salient moments \citep{bordalo2022salience, chetty2009salience}. For MOOC learners, the moment of course completion is highly salient. However, credential sharing is typically encouraged through follow-up communications, reminder emails or platform notifications, that occur after this salient moment has passed. These subsequent prompts might be considerably less attention-grabbing than the completion experience itself.

\subsection{Platform Context: Learning from Prior Experiments}

Prior to this research, Coursera conducted exploratory experiments to understand whether interventions encouraging credential sharing could be effective. We report results from an email campaign encouraging learners who recently graduated to share their certificates on LinkedIn.

Between August 2021 and June 2021, 855 thousand learners who graduated with a course certificate were randomized into treatment and control. The treatment group received an email reminding them to share a certificate and a link to the certificate page with a streamlined sharing feature (clicking redirects to learner's LinkedIn profile page), which reduced sharing friction. Coursera tracked views on the certificate page from LinkedIn (and other business networking platforms) as the main outcome. Receiving a view means that the certificate was present on the LinkedIn profile and that someone clicked on the certificate.\footnote{In addition, Coursera conducted an outcomes survey, which asked about career outcomes; however, less than 1.5\% of subjects respond to the survey.} Appendix \ref{appendix_mailspin} shows summary statistics and covariate balance from this experiment.

\paragraph{Results on credential views.} 

Table \ref{cred_views_mail_spin} reports the estimates of the average treatment effect and conditional average treatment effects by subgroup of the mailing experiment. The outcome is whether or not a certificate received a view. We use a difference-in-means estimator.

\begin{table}[!h]
\caption{Average and Heterogeneous Treatment Effects in Mail Spin Experiment.}\label{cred_views_mail_spin}
\centering
\resizebox{\linewidth}{!}{
\fontsize{10}{12}\selectfont
\begin{tabular}[t]{lccccc}
\toprule
\toprule
\multicolumn{1}{c}{ } & \multicolumn{2}{c}{Panel A: Overall} & \multicolumn{3}{c}{ } \\
\cmidrule(l{3pt}r{3pt}){2-3}
\multicolumn{1}{c}{ } & \multicolumn{2}{c}{Panel B: Characteristic value = 0} & \multicolumn{2}{c}{Panel B: Characteristic value = 1} & \multicolumn{1}{c}{ } \\
\cmidrule(l{3pt}r{3pt}){2-3} \cmidrule(l{3pt}r{3pt}){4-5}
  & Baseline & ATE & Baseline & CATE & Diff. Subgroups\\
\midrule
\addlinespace[0.3em]
\multicolumn{6}{l}{\textbf{Panel A: Overall}}\\
Overall & 0.166 (0.000) & 0.042 (0.001) & -- & -- & --\\
\addlinespace[0.3em]
\multicolumn{6}{l}{\textbf{Panel B: Heterogeneity by Subgroups}}\\
No College Degree & 0.212 (0.000) & 0.026 (0.002) & 0.144 (0.000) & 0.049 (0.001) & 0.023 (0.002)\\
Developing Country & 0.157 (0.000) & 0.014 (0.001) & 0.175 (0.000) & 0.069 (0.001) & 0.055 (0.002)\\
New User  & 0.131 (0.000) & 0.011 (0.001) & 0.180 (0.000) & 0.054 (0.001) & 0.043 (0.002)\\
Tech Domain & 0.135 (0.000) & 0.059 (0.001) & 0.198 (0.000) & 0.024 (0.001) & -0.035 (0.002)\\
\bottomrule
\bottomrule
\end{tabular}
}
\caption*{\footnotesize{\textit{Note: Outcome is Certificate Views - a binary variable taking the value of 1 when the certificate received a view and 0 otherwise. Baseline columns report the mean outcome in the control group. ATE/CATE columns report the difference in means (treated \$-\$ control) with Welch standard errors in parentheses. The overall ATE is shown in Panel A. Panel B reports conditional ATEs by subgroups: No College Degree, Developing Country, New User (below the sample mean of prior course completions), and Tech Domain. The last column reports the difference in CATEs between subgroup value 1 and 0, with its standard error.
}}}
\end{table}

The experiment yielded three key insights. First, the simple mailing campaign was very effective. We estimate that the share of learners who received a view on the certificate increased by 4.2 p.p., which corresponds to an over 25\% increase from the baseline value. Second, the effects were substantially heterogeneous across learner populations. Learners from developing countries experienced a 6.9 p.p.  increase in the probability of receiving a credential view, representing a 39 percent increase from baseline, compared to 1.4 p.p. for learners from developed countries. Similarly, learners without college degrees saw a 4.9 p.p. increase (34 percent from baseline), while college-educated learners experienced only a 2.6 p.p. increase. New learners and learners in non-tech domains saw larger impacts too. Following this experiment, emails with reminders to share credentials became a standard feature of Coursera learners' user experience.

\paragraph{Implications for Our Study Design.}

While the email encouragement experiment demonstrated that light-touch interventions can increase credential visibility, it did not separately identify the effects of sharing behavior versus employer attention, nor did it measure whether increased visibility translates into employment gains. These limitations motivated our follow-up experiment and shaped the research design in two ways. First, the heterogeneity patterns from the email experiment suggested targeting populations with the greatest increases in credential visibility: learners from developing countries and those without college degrees. Second, the measurement challenges highlighted the need for a systematic approach to tracking employment outcomes. We therefore developed a strategy to link Coursera completion records to LinkedIn profiles, allowing us to observe employment outcomes for a subset of learners (see Section \ref{sec:data} for details).

Furthermore, replicating the credential visibility effects across different intervention designs, email campaigns in the earlier study versus in-app notifications in the present study (Section \ref{sec:experiment_design}), provides confidence that the core mechanism operates independently of implementation details. Specifically, if increased visibility to potential employers drives employment effects, these effects should persist across different methods of encouraging credential sharing. This external validity strengthens our interpretation that the employment effects we document arise from the credential signal itself rather than from idiosyncratic features of any particular intervention.

\subsection{Experimental Design}\label{sec:experiment_design}

\subsubsection{Sample Selection and Recruitment}

The experimental sample was restricted to learners from developing countries and learners without college degrees who graduated with credentials in the selected primary domains (Information Technology, Data Science, Computer Science, and Business) between September 2022 and March 2023.\footnote{We follow the OECD classification to categorize countries as developing or developed. The complete list of countries and their classifications is provided in Appendix \ref{dev_countries}.} All Coursera learners meeting these criteria were recruited into the experiment. The experimental population consisted of 880,000 learners, with 37 percent completing Business credentials, 25 percent in Computer Science, 24 percent in Data Science, and 14 percent in Information Technology. The credentials were issued from 7,355 unique courses, ranging from shorter formats such as individual Courses (82 percent) and Guided Projects (16 percent) to longer programs such as Specializations (1.6 percent) and Professional Certificates (1 percent). Appendix \ref{map_world} visualizes the geographical distribution of the experimental sample.

This sample selection differs from Mail Spin's approach of sampling from the entire Coursera user base. By focusing on developing countries and non-college-educated learners, the groups showing the largest credential visibility effects in Mail Spin, we intentionally sample the populations most likely to benefit from the intervention. This design choice reflects both the prediction that credentials should be most valuable for workers with weak alternative signals and the empirical evidence from Coursera's prior experimentation.

\subsubsection{Treatment: The Credential Feature}

In the experiment, the treatment group was randomized to receive access to the \textit{Credential Feature}, a new feature composed of notifications that encouraged the sharing of credentials on LinkedIn and provided a simplified process to do so. This intervention had three key components.

\textbf{Salience and timing.} The first notification was sent on the learner's first visit to the Coursera app after the credential was granted, capturing attention at the moment when completing the course was most salient. The message stated: ``\textit{Do you want to boost your career? Only [XYZ]\% of learners manage to complete [course name] on Coursera and get a certificate. Let everyone know you did it! Add the certificate to your LinkedIn profile in just two clicks}.'' This notification emphasized the achievement and made the call to action concrete and immediate.

If the learner did not click the ``Share now'' button in the first notification, they received a second notification during their subsequent visit to the app, stating: ``\textit{Looking to boost your career? LinkedIn profiles with credentials receive 6x more views! Don't waste your hard-earned certificate! Add the certificate to your LinkedIn profile in just two clicks. PS. This is your last reminder.}'' 

\begin{figure}[H]
    \centering
    \caption{Screenshot of Coursera web app with a notification}
    \includegraphics[scale = 0.7]{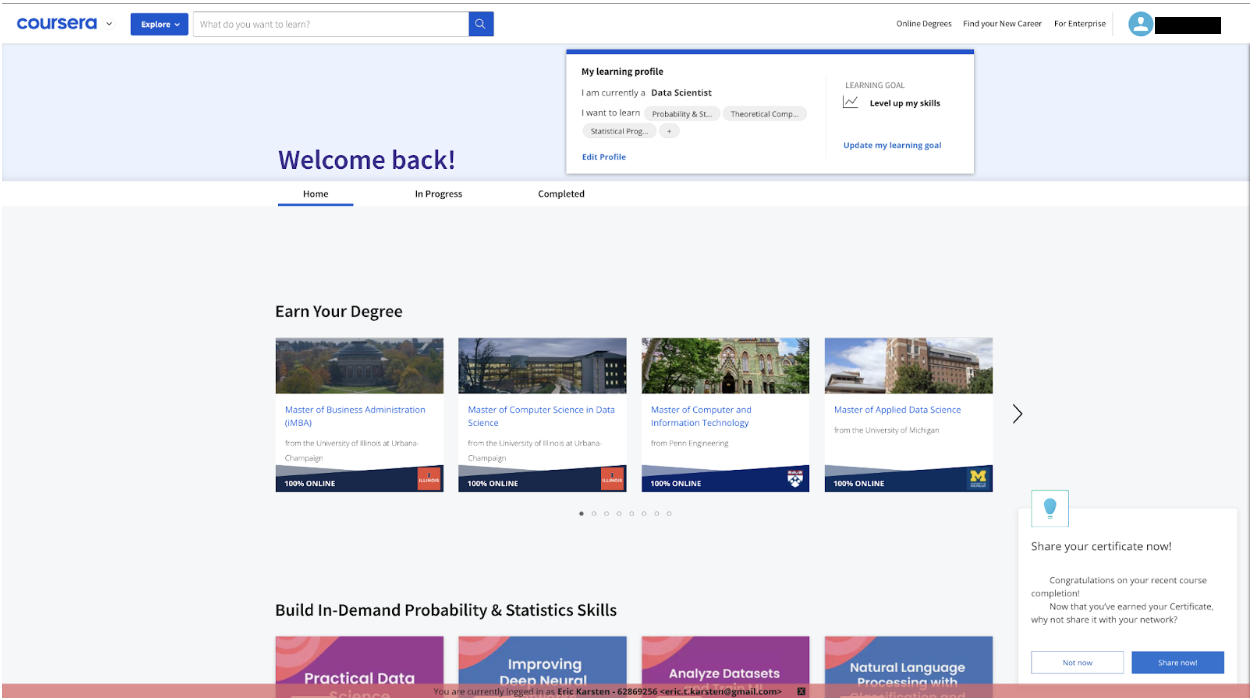}
    \caption*{\footnotesize{\textit{The in-app notification is in the bottom right corner}}}\label{fig:in_app_notification}
\end{figure}

\textbf{Friction reduction.} Both notifications highlighted a streamlined process to add certificates to LinkedIn profiles, which required only two clicks, compared to the baseline case where learners had to manually copy a link from their credential page and paste it into their LinkedIn profile, a process requiring multiple steps across different platforms. Figure \ref{fig:in_app_notification} shows how the notification appeared in the Coursera web application. The control group did not receive these notifications or the streamlined credential-sharing process.

\subsubsection{Randomization Procedure}

Randomization was carried out in monthly batches from September 2022 to March 2023. Each batch included learners who received the certificate between the first and last day of the month. The size of the batches varied from 130,000 to 160,000 learners. At the end of each month, the learners in a given batch were randomized to treatment and control groups. Randomization was stratified based on the learners' primary domain (Business, Computer Science, Data Science, or Information Technology), whether they came from a developing country, and whether they had a college degree. Each learner was randomized to the treatment or control group with equal probability within these strata, resulting in approximately 50 percent assignment to each group.

This strata-based randomization ensured balance across key dimensions that could affect both credential-sharing behavior and employment outcomes. Each batch can be viewed as a mini-experiment, and we verify that results are consistent across batches (with one notable exception in Batch 5 where the estimates of the average treatment effect are positive but statistically insignificant). Appendix \ref{appendix_batches} provides details of the batched recruitment.

Appendix \ref{balance_appendix} presents detailed balance checks comparing treatment and control groups on pre-treatment covariates. We do not find statistically significant differences in covariate values between the treatment and control groups.

\subsubsection{Intention-to-Treat Design and Exposure}

Learners who did not launch the Coursera app within 30 days of being assigned to the treatment group, did not receive any notification, as the intervention was delivered through in-app messages. However, we continue to consider these users to be part of the treatment group, following standard intention-to-treat (ITT) framework. This design choice makes our estimates conservative: they represent the effect of offering access to the Credential Feature, not the effect of actually seeing the notification.

The percentage of learners who launched the app after graduation varied across batches, with the second batch having the highest percentage (96 percent of learners) and the fifth batch having the lowest (82 percent). This variation in exposure rates contributes to heterogeneity in treatment effects across batches. Note that the batch is defined based on graduation date rather than app visit dates. Thus, a learner who graduated in September (the first batch) but visited the Coursera app and saw the notification in October is still classified as Batch 1. 

\subsection{Data and Outcome Measurement}\label{sec:data}

To analyze the results of the experiment, we combined data from two sources, allowing us to measure outcomes at different stages of the causal chain from intervention to employment.

\subsubsection{Coursera Internal Data (N=880,000)}

The first data source is \textit{Coursera Internal Data}, which includes platform records describing user engagement with Coursera apps and responses to user registration surveys. For each learner, we observe the name and dates of granted certificates, enabling us to precisely track the timing of course completion and randomization.

\textbf{Certificate engagement metrics.} For each credential, we observe the level of engagement on the certificate pages, including the number of views (page visits) on the certificate page hosted on Coursera's website. Each view is categorized by its origin, including whether the view came from LinkedIn (the referral page was LinkedIn) and whether the view came from the learner themselves. The latter metric is inferred by Coursera from several signals (IP address, login status, timing patterns) and may contain both Type I and Type II errors; we therefore present results both including and excluding own views. These view metrics serve as intermediate outcomes in the causal chain: if the treatment increases credential sharing, and shared credentials are visible on LinkedIn, we should observe increased views from LinkedIn users (potentially including recruiters and employers).

\textbf{Skill and credential characteristics.} Each certificate is associated with a primary domain (Business, Computer Science, Data Science, or Information Technology) and specific skills such as ``project management'', ``digital marketing'', or ``web development''. In our dataset, we observe 462 different skills across the 7,355 courses. For each learner, Coursera assesses skill mastery using a Glicko-based algorithm \citep{reddick2019using} and assigns a score reflecting their demonstrated proficiency. We use both the absolute skill score and a max-mean standardization computed within each specific skill category to capture learner ability.\footnote{The standardized skill score is calculated as: (learner's skill score - mean skill score)/(max skill score), where mean and max are computed at the level of each specific skill. This standardization allows comparison across skills with different scoring scales.}

\textbf{Demographic and background characteristics.} We observe the country where the learner registered for the course. Following the OECD classification, we use this information to group countries into developing and developed. Additionally, learners voluntarily provided information about their level of education and gender through the registration survey. Because response to the survey is voluntary, we do not observe these characteristics for all learners (64.5 percent did not report education level; 53.3 percent did not report gender in the full sample).

\subsubsection{LinkedIn Matched Sample (N=37,000)}\label{linkedin_sample}
The second and primary data source for measuring employment outcomes is the \textit{LinkedIn Matched Sample}. Upon enrolling in courses, learners were asked to provide their LinkedIn profile URLs. This request was part of Coursera's standard enrollment process and occurred before any awareness of the credential-sharing experiment. 

In September 2023, we collected publicly available information from the LinkedIn profiles of approximately 37,000 learners who had provided LinkedIn URLs and for whom profiles were still accessible.\footnote{Some profiles became inaccessible due to account deletion, privacy setting changes, or URL errors. Our final LinkedIn Matched Sample represents approximately 90 percent of learners who initially provided URLs.} The timing of data collection resulted in a 12-month gap between randomization and LinkedIn data collection for learners in the first batch (September 2022 graduates) and an 8-month gap for those in the last batch (March 2023 graduates).

\textbf{Employment history.} From LinkedIn profiles, we extracted detailed employment histories including job titles, employer names, start dates, and (when reported) end dates for all positions listed on the profile. This information allows us to identify new employment reported after the intervention and to classify jobs according to their relevance to the credential earned.

\textbf{Educational and experience background.} The LinkedIn data includes information on educational background (degrees earned, institutions attended, graduation years), work experience (number of positions held, total years of experience, industry sectors), and LinkedIn activity metrics (profile completeness, number of connections when available). These variables allow us to construct rich measures of baseline employability and alternative signals available to the learner. Details about the construction of features using LinkedIn data are provided in Appendix \ref{app:linkedin_features}.

\textbf{Resume construction for counterfactual scoring.} The comprehensive LinkedIn profile data described above serves an additional purpose in our mechanism analysis. For the resume scoring exercise in Section \ref{sec:resume_scoring}, we construct complete resume representations for each learner using all available profile information—including educational background, employment history, skills, and other professional details. We create counterfactual resumes by comparing profiles with and without the earned Coursera credential, holding all other elements constant. This approach allows us to isolate the marginal effect of credential visibility on perceived candidate quality while preserving the realistic context of each learner's actual professional background.

\textbf{Selection into the LinkedIn Matched Sample.} Comparing the average covariate values in \textit{Coursera Internal Data} and the \textit{LinkedIn Matched Sample} reveals that learners who provided LinkedIn URLs differ systematically from the broader experimental population. LinkedIn-matched learners are more likely to have graduated with a certificate in Data Science (30.0 percent versus 23.6 percent, difference 6.4pp with S.E. <0.001), less likely to have participated in a Guided Project (9.7 percent versus 16.8 percent, difference -7.1pp with S.E. 0.002), and slightly less likely to be from a developing country (85.0 percent versus 89.6 percent, difference -4.6pp with S.E. 0.002). 

Learners in the LinkedIn Matched Sample are also more likely to report demographic information in the Coursera registration survey. The share reporting gender is 89.9 percent in the LinkedIn sample versus 46.7 percent in the full sample; the share reporting education level is 83.6 percent versus 35.5 percent. This pattern suggests that LinkedIn-matched learners are more engaged with professional networking and career development activities, which likely correlates with baseline employability and job search intensity.

While this selection limits the external validity of our employment results to the subset of MOOC learners with LinkedIn profiles, it does not threaten the internal validity of our experimental estimates. Randomization occurred before LinkedIn URL provision, and balance checks (reported in Appendix \ref{balance_appendix}) confirm no significant differences in observable characteristics between treatment and control groups within the LinkedIn Matched Sample. Moreover, we can assess the generalizability of our findings by examining certificate view outcomes in the full experimental sample, which we present in Section \ref{sec:views}.

\subsubsection{Outcome Definitions}

We define outcomes at three stages of the hypothesized causal chain from intervention to employment: credential sharing, credential visibility, and employment.

\textbf{Primary outcomes: Employment (LinkedIn Matched Sample only).} 

\emph{New Job} is an indicator equal to one if the learner reports a new position on LinkedIn with a starting date at least one month after randomization, and zero otherwise.\footnote{In robustness checks, we alternatively require jobs to start at least four months after randomization to address concerns that treatment might prompt learners to update their profiles with jobs found before the intervention but not yet reported on LinkedIn.} This broad definition includes all new positions, whether with new employers or internal promotions, and regardless of sector or skill relevance.

\emph{New Job in Scope} is an indicator equal to one if the learner reports a new position in the technology sector or a managerial role with a starting date at least one month after randomization. We identify relevant positions by searching job titles for keywords including \emph{software}, \emph{data}, \emph{developer}, \emph{engineer}, \emph{analyst}, \emph{manager}, and related terms. This restriction aims to capture jobs where the skills taught in the MOOC credential are likely to be directly applicable, providing a more conservative test of the credential's labor market value.

\emph{Credential Shared} is an indicator equal to one if the credential appears on the learner's LinkedIn profile at the time of data collection (September 2023) and zero otherwise.\footnote{Credentials are identified by their unique Coursera-issued ID, which is displayed next to the credential's name on LinkedIn profiles, allowing precise matching.} This variable serves as the first-stage outcome when evaluating the Credential Feature intervention and as the treatment variable when estimating the causal effect of credential sharing on employment using the encouragement design framework.

\textbf{Secondary outcomes: Credential visibility (Full sample).}

For all learners in \textit{Coursera Internal Data}, we observe the number of visits to the credential page hosted on Coursera's website and the referring page for each visit. Based on this data, we construct four indicator variables measuring credential visibility. \emph{All Views} equals one when the certificate received at least one view. \emph{All Views by Others} equals one when the certificate page was viewed at least once by someone other than the learner (excluding views Coursera classifies as originating from the learner's IP address or login session). \emph{Views LinkedIn} equals one when at least one view originated from LinkedIn (referral URL indicates LinkedIn). \emph{Views LinkedIn by Others} equals one when at least one view originated from LinkedIn and was made by someone other than the learner.

These visibility metrics provide two important pieces of information. First, they allow us to examine treatment effects in the full experimental sample (N=880,000), not just the LinkedIn Matched Sample. Second, they serve as intermediate outcomes in the causal chain: if shared credentials lead to employment by increasing visibility to employers, we should observe that treatment increases views and that views predict employment outcomes.

\subsubsection{Summary Statistics}

Table \ref{pretreatment_sum_stats} presents summary statistics for both the full Coursera Internal Data sample and the LinkedIn Matched Sample. Panel A reports pre-treatment covariates, while Panel B reports outcome variables. 

\begin{table}[!htbp] \centering 
\caption{Summary statistics pretreatment and outcome variables}\label{pretreatment_sum_stats}
  \resizebox{0.8\textwidth}{!}{
        \begin{tabular}{@{\extracolsep{5pt}}lcc|cc}
        \\[-1.8ex]\hline
        \hline \\[-1.8ex]
        & \multicolumn{2}{c}{Coursera Internal Data} & \multicolumn{2}{c}{LinkedIn Matched Sample} \\
        Variable name & \multicolumn{1}{c}{Mean} & \multicolumn{1}{c}{S.E.} & \multicolumn{1}{c}{Mean} & \multicolumn{1}{c}{S.E.} \\
        \hline \\[-1.8ex] 
        Treatment & 0.499 & 0.001 & 0.500 & 0.003 \\
        \midrule
        \multicolumn{5}{@{}l}{\textit{Panel A: Pre-treatment covariates}}\\
        \addlinespace
        Professional Experience Years & -- & -- & 3.040 & 0.028 \\
        Past Tech Job & -- & -- & 0.127 & 0.002 \\
        Past Managerial Job & -- & -- & 0.064 & 0.001 \\
        Main Skill Absolute & 0.099 & 0.001 & 2.074 & 0.010 \\
        Main Skill Standardized & 0.000 & $<$0.001 & 0.000 & 0.001 \\
        Computer Science & 0.252 & 0.001 & 0.230 & 0.002 \\
        Data Science & 0.236 & 0.001 & 0.300 & 0.002 \\
        Information Technology & 0.140 & 0.001 & 0.138 & 0.002 \\
        Guided Project & 0.168 & 0.001 & 0.097 & 0.002 \\
        Professional Certificate & 0.005 &  $<$0.001 & 0.005 &  $<$0.001 \\
        Specialization & 0.009 &  $<$0.001 & 0.009 & 0.001 \\
        Developing Country & 0.896 & 0.001 & 0.850 & 0.002 \\
        Associate Degree & 0.029 &  $<$0.001 & 0.062 & 0.001 \\
        Bachelor Degree & 0.127 & 0.001 & 0.367 & 0.003 \\
        Some College & 0.072 & 0.001 & 0.130 & 0.002 \\
        Doctorate Degree & 0.004 &  $<$0.001 & 0.012 & 0.001 \\
        High School Diploma & 0.059 & 0.001 & 0.097 & 0.002 \\
        Less than High School & 0.009 &  $<$0.001 & 0.012 & 0.001 \\
        Masters Degree & 0.050 & 0.001 & 0.146 & 0.002 \\
        No Education Mentioned & 0.645 & 0.002 & 0.164 & 0.002 \\
        Professional Degree & 0.004 &  $<$0.001 & 0.010 & 0.001 \\
        Male & 0.302 & 0.002 & 0.674 & 0.002 \\
        Gender Not Mentioned & 0.533 & 0.002 & 0.101 & 0.002 \\
        \midrule
        \multicolumn{5}{@{}l}{\textit{Panel B: Outcome variables}}\\
        \addlinespace
        New Job & -- & -- & 0.177 & 0.002 \\
        New Job in Scope & -- & -- & 0.133 & 0.002 \\
        Credential Shared & -- & -- & 0.181 & 0.002 \\
        All Views & 0.191 & 0.001 & 0.429 & 0.003 \\
        All Views by Others & 0.143 & 0.001 & 0.318 & 0.002 \\
        Views LinkedIn & 0.165 & 0.001 & 0.409 & 0.003 \\
        Views LinkedIn by Others & 0.124 & 0.001 & 0.296 & 0.002 \\
        \\[-1.8ex]\hline
        \hline \\[-1.8ex]
        \end{tabular}
    }
\caption*{\footnotesize{\textit{Note: Professional Experience Years is the number of years between the starting date of the first job listed on LinkedIn and August 2023. Past Tech Job takes the value of 1 when the learner had a job title related to technology before randomization and zero otherwise. Past Managerial Job is defined analogously for jobs with managerial titles.}}}
\end{table}

In the \textit{LinkedIn Matched Sample}, 17.7 percent of learners reported new jobs during the follow-up period (8-12 months depending on batch), and 13.3 percent reported jobs in scope, indicating that most new jobs were related to the certificate domains. The baseline credential-sharing rate was 18.1 percent, confirming that even among LinkedIn users, a selected sample likely more aware of the value of credential sharing—the majority of learners did not add their credentials without intervention.

Certificate view rates were substantially higher in the LinkedIn Matched Sample compared to the full sample, consistent with LinkedIn-matched learners being more engaged with professional networking. In the LinkedIn Matched Sample, 42.9 percent of credentials received at least one view, and 29.6 percent received views from LinkedIn by others. In contrast, in the full \textit{Coursera Internal Data}, only 19.1 percent of credentials received any views, and 12.4 percent received views from LinkedIn by others. This difference likely reflects both higher credential-sharing rates among LinkedIn users and greater overall engagement with the platform ecosystem.

%% file: 4.main_result.tex
\section{Results}
\label{sec:results}

This section presents the results of our randomized experiment on the labor market effects of credential-sharing interventions. We focus primarily on the \emph{LinkedIn Matched Sample}, where we observe both credential-sharing behavior and employment outcomes. This allows us to estimate the intervention's effect on obtaining new employment and to separately identify whether credentials are shared and whether shared credentials attract employer attention.

We organize the analysis as follows. Section \ref{sec:intent_to_treat} presents intent-to-treat estimates of the \emph{Credential Feature} on both credential sharing and employment outcomes, establishing the average causal effect of the intervention. Section \ref{sec:hte} examines heterogeneous treatment effects, showing that the intervention is most effective for learners with low baseline employability and that those most responsive to the treatment are also those who benefit most from credential sharing. Section \ref{sec:iv} uses the randomized treatment as an instrument to estimate the local average treatment effect of credential sharing itself on employment outcomes for compliers. Finally, Section \ref{sec:views} examines credential visibility in the full Coursera sample.

\subsection{Intent-to-Treat Effects on Credential Sharing and Employment}
\label{sec:intent_to_treat}

We begin by examining the average causal effect of the \textit{Credential Feature}, the intervention that streamlined credential sharing and provided an in-app notification prompting learners to add their course completion to LinkedIn. This intent-to-treat analysis captures the combined effect of encouragement to share and reduced friction in the sharing process, corresponding to the overall benefit of deploying the feature to all Coursera learners.

\subsubsection{Evolution of Employment Outcomes}

\begin{figure}[htp]
\caption{Share of learners reporting new jobs in treatment and control groups}\label{fig:outcomes_per_batch}
\centering
\includegraphics[scale = 0.6]{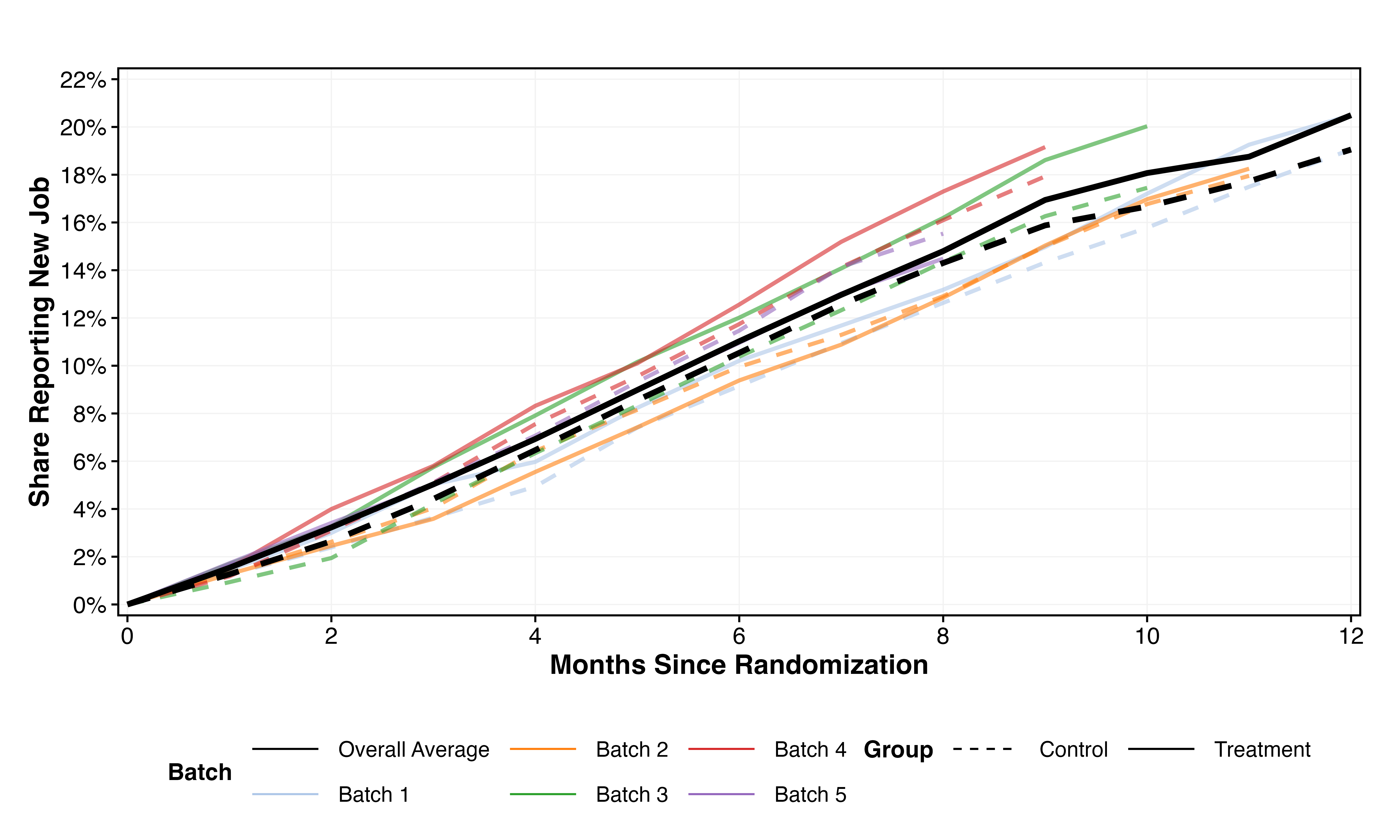}
\caption*{\footnotesize{\textit{Note: For each batch, the figure presents the share of learners by treatment and control groups who have reported a new job. Solid lines present the treatment groups, dashed lines the control groups. The thick black line shows the average outcome in an experimental group across all batches. }}}
\end{figure}

Figure \ref{fig:outcomes_per_batch} displays the share of learners reporting new jobs over time, separately by treatment status and experimental batch. Month 0 corresponds to the month of randomization, which differs across batches in calendar time. Several patterns emerge. First, treated and control groups exhibit similar trends, with approximately 10\% of learners reporting new jobs after six months across all batches. Second, in all batches except Batch 5, a larger share of treated learners report new jobs compared to control learners. The divergence between treatment and control emerges gradually rather than immediately, consistent with job search and hiring processes taking time.

\subsubsection{Average Treatment Effects}

Table \ref{combined_table} presents estimates of the average treatment effect using Cox proportional hazards models that account for differential follow-up periods across batches. Column 1 shows that treatment increases the probability of sharing credentials by 2.8 percentage points (standard error 0.4 percentage points), representing a 17\% increase from the control group baseline of 16.4\%. This substantial first-stage effect establishes that the intervention successfully reduced barriers to credential sharing.

Columns 2 and 3 examine effects on employment for jobs reported with a start date at least one month after randomization. For the \textit{New Job} outcome, which includes any new position reported on LinkedIn (Column 2), we estimate a treatment effect of 1.01 percentage points (standard error 0.45 percentage points), representing a 5.8\% increase relative to the control group baseline of 17.3\%. The effect is larger when restricting to \textit{New Job in Scope} (Column 3)---positions in professional, technical, or managerial roles likely to value MOOC credentials---with a treatment effect of 1.19 percentage points (standard error 0.40 percentage points), or a 9.3\% increase from the baseline of 12.7\%.

\begin{table}[!htbp]
\centering
\caption{Intent-to-Treat Effects on Credential Sharing and Employment}
\label{combined_table}
\begin{tabular}{@{}lcccccc@{}}
\toprule
\toprule
 & \multicolumn{2}{c}{ATE} & \multicolumn{2}{c}{ATE (\%)} & Baseline (\%) & $N$ \\
\cmidrule(lr){2-3}\cmidrule(lr){4-5}
 & (pp) & SE (pp) & (\%) & SE (\%) &  &  \\
\midrule
Credential Shared & 2.887 & 0.400 & 17.3 & 2.4 & 16.7 & 36,946 \\
New Job & 1.020 & 0.452 & 5.9 & 2.6 & 17.3 & 36,946 \\
New Job in Scope & 1.198 & 0.397 & 9.4 & 3.1 & 12.7 & 36,946 \\
New Job ($\geq$4 months) & 0.620 & 0.329 & 6.9 & 3.6 & 9.0 & 36,946 \\
New Job in Scope ($\geq$4 months) & 0.673 & 0.275 & 11.4 & 4.7 & 5.9 & 36,946 \\
\midrule
\multicolumn{7}{l}{\textit{Panel B: Restricted Sample (Prior Job Not in Scope)}} \\
\midrule
New Job in Scope & 0.916 & 0.363 & 10.6 & 4.2 & 8.7 & 30,607 \\
New Job in Scope ($\geq$4 months) & 0.626 & 0.353 & 7.2 & 4.1 & 8.7 & 30,607 \\
\bottomrule
\bottomrule
\end{tabular}
\caption*{\footnotesize{\textit{Note: This table presents intent-to-treat estimates from Cox proportional hazards models (for employment outcomes) and OLS (for credential sharing). Panel A shows results for the full LinkedIn matched sample. Panel B restricts to learners whose previous employment was not in scope. ATE (pp) reports the treatment effect in percentage points. ATE (\%) reports the percentage change relative to the control group baseline. Baseline (\%) is the outcome rate in the control group. Jobs starting at least one month after randomization are included in rows 2--3; rows 4--5 restrict to jobs starting at least four months post-randomization. All models include controls for batch, education, gender, domain, credential type, country, professional experience, and prior employment in technical and business roles.}}}
\end{table}

\paragraph{Addressing Profile Updating Concerns.}

A potential concern is that treated learners might simply update their LinkedIn profiles to add jobs they had already obtained but not yet reported, rather than the intervention leading to new employment. We address this concern in two ways.

First, Columns 4 and 5 restrict attention to jobs reported with a start date at least four months after randomization. At the time of treatment, it is implausible that learners would already know about jobs starting several months later. The treatment effects persist when imposing this restriction. For \textit{New Job}, we estimate a 0.61 percentage point increase (standard error 0.33 percentage points), or 6.8\% from the baseline of 9.0\% (Column 4). For \textit{New Job in Scope}, the effect is 0.67 percentage points (standard error 0.27 percentage points), representing an 11.3\% increase (Column 5). These results indicate that the intervention generates employment gains rather than accelerates the timing of profile updates.

Second, we investigate whether the intervention prompted treated learners to become more active on LinkedIn generally, which could confound the interpretation of employment effects. We measure profile completeness by counting the total number of characters across all fields in each learner's LinkedIn profile, excluding fields related to credentials and new jobs. Treated learners have profiles averaging 512 characters compared to 511 characters for control group learners, a difference of 1 character that is statistically insignificant ($t = -0.23$, $p = 0.82$). A Kolmogorov-Smirnov test confirms that the distributions of profile lengths are statistically indistinguishable between treatment and control groups ($D = 0.009$, $p = 0.80$). Appendix \ref{appendix_profile} displays the full distributions, showing high overlap between treatment and control groups. These results confirm that the employment effects operate through credential presence rather than general improvements in the richness of the LinkedIn profile quality.

\subsection{Heterogeneous Treatment Effects I: Baseline Employability}\label{sec:hte}

In this section, we examine heterogeneity along two dimensions: baseline employability and treatment responsiveness.
\subsubsection{Baseline Employability}
\paragraph{Predicting Baseline Employability.}

Following \citet{athey2023machine}, we use a gradient boosting machine (GBM) model to predict each learner's baseline probability of finding employment absent treatment. We train the model on the control group, predicting the new job as a function of pre-treatment characteristics including demographic information, educational background, prior employment, course enrollment patterns, platform engagement, and skill assessments.\footnote{We use 5-fold cross-validation to tune hyperparameters and avoid overfitting.} We then apply the trained model to both treatment and control groups and partition learners into terciles based on predicted employability: low (bottom tercile), medium (middle tercile), and high (top tercile). Table \ref{tertiles_employ} shows that baseline employability varies across groups. The high employability tercile has a 29\% employment rate in the baseline, while the medium and low terciles have 13\% and 11\% rates respectively.

\paragraph{Treatment Effects by Baseline Employability.} Table \ref{tertiles_employ} presents heterogeneous treatment effects across employability terciles using Cox proportional hazards models. For the \emph{new job} outcome, treatment effects are small and statistically insignificant for high employability learners (0.3 percentage points) but larger and significant for medium and low employability learners (1.2 and 1.4 percentage points respectively). The pattern is similar but more pronounced for \emph{new job in scope}, with effects of 0.9 percentage points for high employability learners and 1.7 and 1.1 percentage points for medium and low employability learners.

\begin{table}[!htbp]
\centering
\caption{Heterogeneous Treatment Effects by Baseline Employability}
\label{tertiles_employ}
\resizebox{0.95\textwidth}{!}{%
\begin{tabular}{@{}lcccccccccc@{}}
\toprule
\toprule
 & \multicolumn{3}{c}{High Employability} & \multicolumn{3}{c}{Medium Employability} & \multicolumn{3}{c}{Low Employability} \\
\cmidrule(lr){2-4} \cmidrule(lr){5-7} \cmidrule(lr){8-10}
 & Baseline & ATE & ATE & Baseline & ATE & ATE & Baseline & ATE & ATE \\
 & (\%) & (pp) & (\%) & (\%) & (pp) & (\%) & (\%) & (pp) & (\%) \\
\midrule
\multicolumn{10}{@{}l}{\textit{Outcome Variable: New Job}} \\
\addlinespace[0.1cm]
Treatment Effect & 29.1 & 0.344 & 1.2 & 12.6 & 1.248 & 9.9 & 11.1 & 1.386 & 12.5 \\
 & (0.6) & (0.994) & (3.4) & (0.5) & (0.756) & (6.0) & (0.4) & (0.600) & (5.4) \\
 \multicolumn{10}{@{}l}{\textit{Outcome Variable: New Job in Scope}}\\
\addlinespace[0.1cm]
Treatment Effect & 21.9 & 0.912 & 4.2 & 8.9 & 1.706 & 19.1 & 8.1 & 1.102 & 13.7 \\
 & (0.5) & (0.881) & (4.0) & (0.4) & (0.675) & (7.6) & (0.3) & (0.516) & (6.4) \\
\addlinespace[0.1cm]
\midrule
Observations & 11,955 & 11,955 & 11,955 & 10,249 & 10,249 & 10,249 & 14,742 & 14,742 & 14,742 \\
\bottomrule
\bottomrule
\end{tabular}
}
\caption*{\footnotesize{\textit{Note: Heterogeneous treatment effects across terciles of predicted baseline employability. Baseline (\%) reports the employment rate in the control group for each tercile. ATE (pp) reports the average treatment effect in percentage points, estimated using Cox proportional hazards models. Standard errors in parentheses.}}}
\end{table}

\begin{figure}
\centering
\caption{Treatment Effects Across the Employability Distribution}
\includegraphics[scale = 0.4]{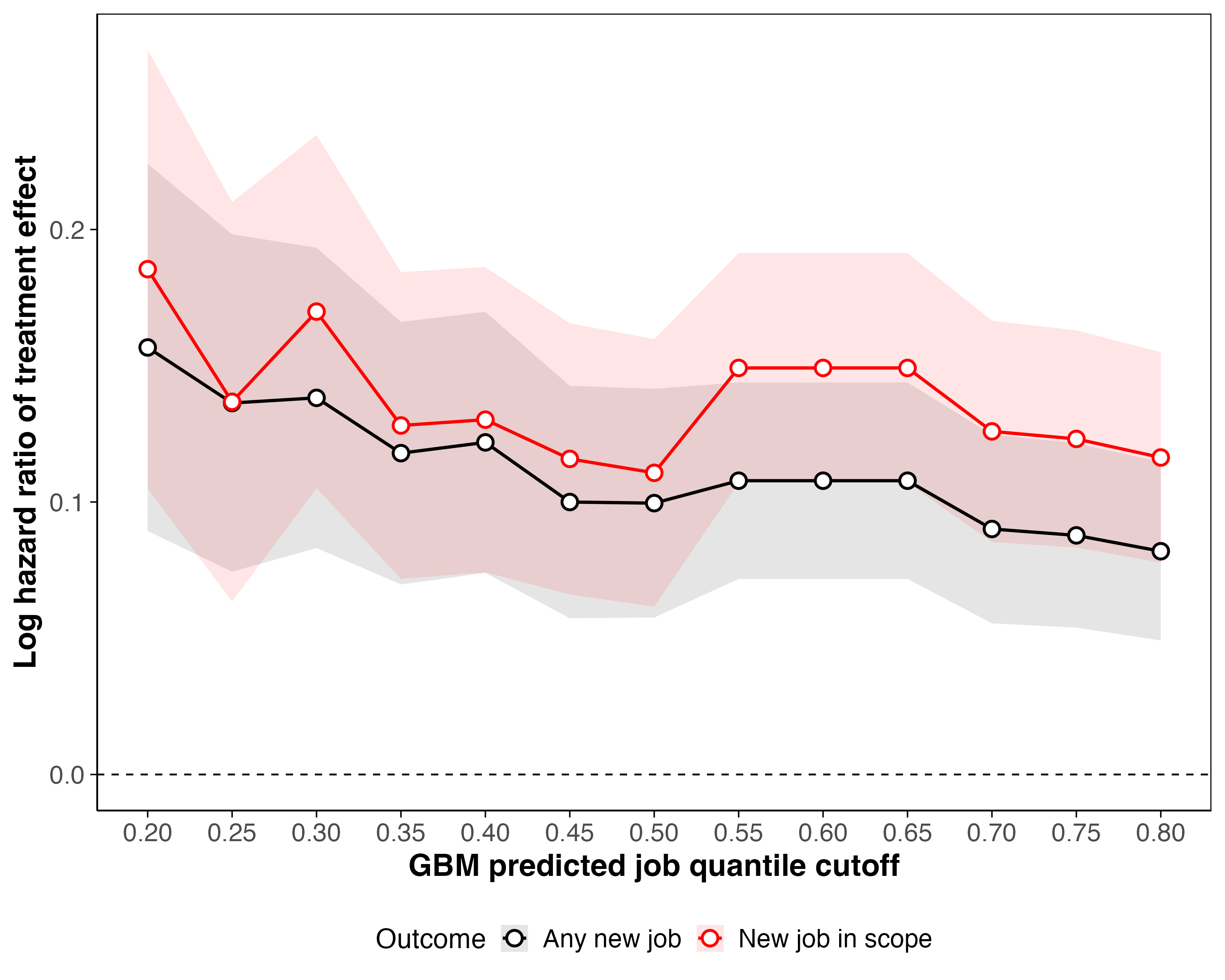}
\label{fig:continuous_het}
\caption*{\footnotesize{\textit{Note: Log hazard ratios from Cox proportional hazards models estimated at cumulative quantile cutoffs of predicted baseline employability. Each point represents the treatment effect for learners with predicted employability at or below the specified percentile threshold. Shaded regions indicate 95\% confidence intervals.}}}
\end{figure}

To explore this heterogeneity more granularly, Figure \ref{fig:continuous_het} traces treatment effects across the full employability distribution. We estimate Cox proportional hazards models at 12 quantile cutoffs ranging from the 20th to 80th percentiles. For each threshold $\tau$, we restrict the sample to learners with predicted employability at or below the $\tau$th percentile and estimate the treatment effect for both outcomes.

We find that the credential intervention primarily benefits low and medium employability learners. Treatment effects are strongest among the least employable learners, with log hazard ratios of 0.16 for any new job and 0.19 for new jobs in scope at the 20th percentile. As we incorporate progressively more employable learners, effects decline monotonically to approximately 0.09 and 0.13 respectively by the 75th percentile. Notably, medium employability learners show particularly strong gains for in-scope employment: the uptick in treatment effects around the 55th-65th percentiles indicates that these learners especially benefit from obtaining jobs aligned with the credential. 

Note that even our high employability tercile has only a 29\% employment rate over the follow-up period, reflecting the sample's focus on learners without college degrees from developing countries.

\subsubsection{Response Heterogeneity}

Previously, we examined heterogeneity based on baseline need, who has the lowest employment prospects absent treatment. We now examine a distinct dimension: treatment responsiveness. Who is most likely to respond to the intervention by sharing credentials, and do those who respond also benefit most? This question is important for efficient targeting: if platforms can identify learners who both respond to nudges and benefit from credential sharing, they can direct interventions toward high-value populations without complex machine learning systems.

We begin by establishing that there is heterogeneity in the treatment effect on the credential sharing outcome. We follow the methodology developed by \citet{yadlowsky2025evaluating}. We randomly split the \emph{LinkedIn Matched Sample} into a training set (one-third of observations) and a test set (two-thirds). Using the training set, we estimate individual-level treatment effects on credential sharing using a Generalized Random Forest (GRF) \citep{wager2018estimation}. We then apply the trained model to the test set to obtain predicted treatment effects for each learner. We define high-RATE learners as those with predicted treatment effect above the median and low-RATE learners as those below the median.

\begin{figure}[!htbp]
\centering
\caption{Targeting Operator Characteristic Curve for Credential Sharing}
\includegraphics[width=0.7\textwidth]{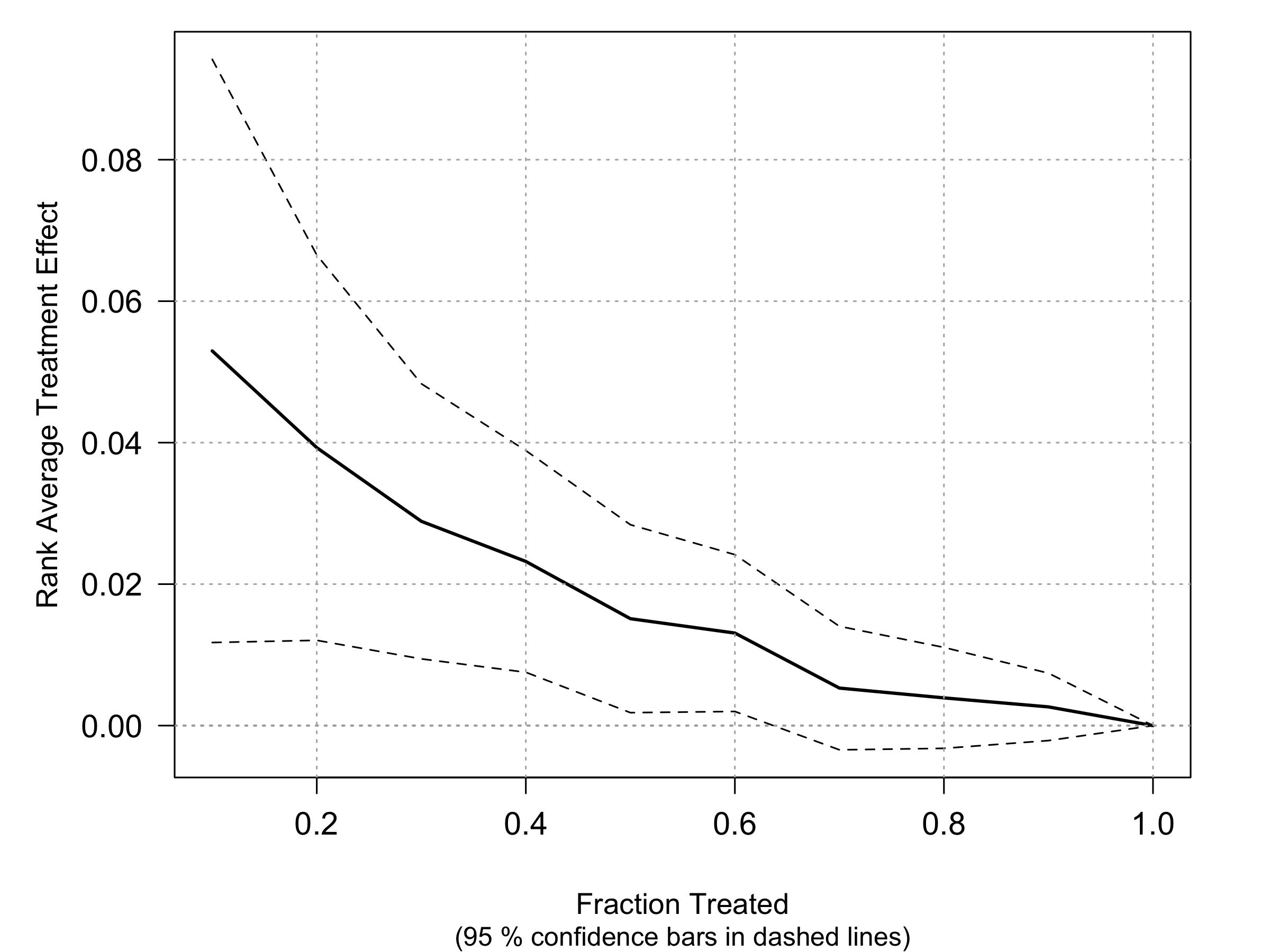}
\label{fig:toc_curve}
\caption*{\footnotesize{\textit{Note: This figure displays the Targeting Operator Characteristic (TOC) curve, showing the average treatment effect on credential sharing as a function of the fraction of the population targeted, ordered by predicted treatment effects from a Generalized Random Forest model. The x-axis shows the fraction of learners treated; the y-axis shows the average treatment effect in the treated fraction. The solid line shows the point estimate; dashed lines present the 95\% confidence interval. Targeting the most responsive learners yields substantially larger treatment effects than uniform targeting.}}}
\end{figure}

This approach has two advantages. First, the forests used for estimating treatment effects and for assessing treatment effect heterogeneity are trained on different datasets. Second, GRF allows for flexible nonlinear interactions between treatment and covariates without imposing parametric restrictions. 
Figure \ref{fig:toc_curve} presents a Targeting Operator Characteristic (TOC) curve, which plots the average treatment effect on credential sharing as a function of the fraction of the population targeted, when targeting is based on predicted treatment effects. The curve reveals substantial heterogeneity in treatment responsiveness. Targeting the top 20\% of predicted responders yields an average first-stage treatment effect 50\% larger than targeting the top 40\%, indicating that observable characteristics strongly predict who will respond to the intervention.

\begin{figure}[!htbp]
\centering
\caption{Characteristics of High-RATE versus Low-RATE Learners}\label{fig:smd_rate}
\includegraphics[width=0.6\textwidth]{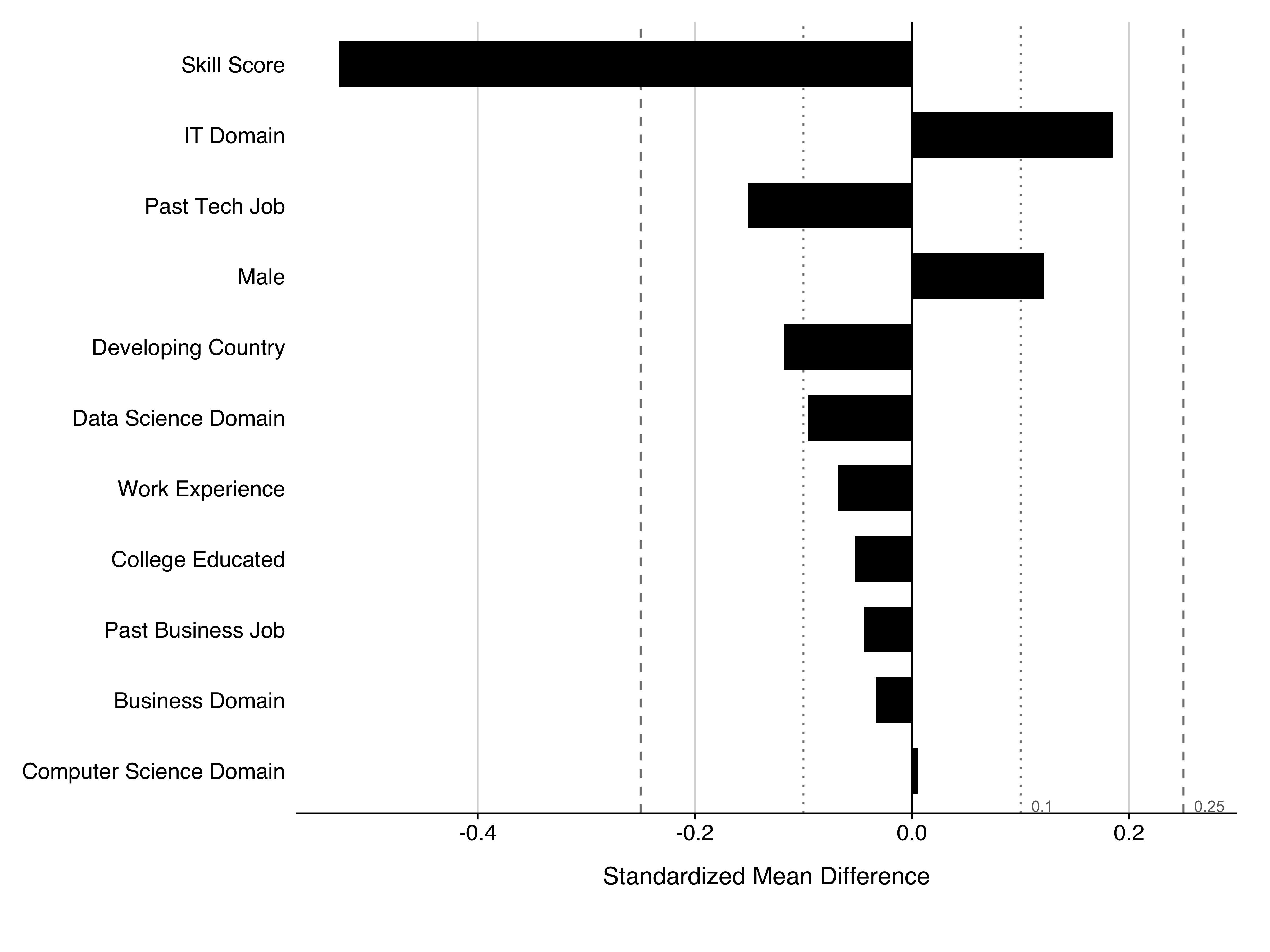}
\caption*{\footnotesize{\textit{Note: This figure displays standardized mean differences in pre-treatment characteristics between high-RATE (above median predicted treatment effect) and low-RATE (below median) groups. High-RATE learners are those predicted to be most responsive to the credential-sharing intervention based on a Generalized Random Forest model. Positive values indicate characteristics more common among high-RATE learners; negative values indicate characteristics more common among low-RATE learners. The largest differences are in skill assessment scores, IT domain enrollment, gender, and prior technical employment.}}}
\end{figure}

Who are the high-RATE learners? Figure \ref{fig:smd_rate} displays standardized mean differences in pre-treatment characteristics between high-RATE and low-RATE groups. Several patterns emerge. High-RATE learners, those most responsive to the treatment, have substantially lower skill assessment scores (standardized mean difference of $-0.4$, the largest difference across all covariates). They are more likely to be from developing countries (SMD of $+0.15$), more likely to be male (SMD of $+0.25$), and more likely to have taken courses in IT domains (SMD of $+0.3$). Conversely, high-RATE learners are less likely to have prior employment in technical roles (SMD of $-0.2$). Educational attainment shows no strong pattern, suggesting that treatment responsiveness is not simply about formal credentials.

\paragraph{Do Responders Benefit More?}

Having characterized who responds to the treatment, we now examine whether treatment responsiveness predicts employment returns to credential sharing. Table \ref{tab:rate_iv} presents instrumental variables estimates separately for high-RATE and low-RATE groups. For each group, we instrument credential sharing with treatment assignment and estimate the effect of sharing on employment.

The results demonstrate that treatment responsiveness strongly predicts employment benefits. For high-RATE learners (Columns 1-2), credential sharing increases employment probability by 10.5 percentage points (standard error 2.8 percentage points) without controls and 10.2 percentage points (standard error 2.9 percentage points) with full controls. Both estimates are statistically significant at the 1\% level. In contrast, for low-RATE learners (Columns 3-4), the estimated effects are 4.2 percentage points (standard error 3.9 percentage points) without controls and 4.0 percentage points (standard error 3.9 percentage points) with controls. Neither estimate is statistically significant. We find that those most responsive to credential-sharing nudges are also those who benefit most from sharing credentials. 

\begin{table}[!htbp]
\centering
\caption{Returns to Credential Sharing by Treatment Responsiveness}
\label{tab:rate_iv}
\resizebox{0.5\textwidth}{!}{%
\begin{tabular}{@{}lcccc@{}}
\toprule
\toprule
 & \multicolumn{2}{c}{High-RATE Group} & \multicolumn{2}{c}{Low-RATE Group} \\
\cmidrule(lr){2-3} \cmidrule(lr){4-5}
 & (1) & (2) & (3) & (4) \\
\midrule
\multicolumn{5}{@{}l}{\textit{Second Stage: Dependent Variable = New Job}} \\
\addlinespace[0.1cm]
Credential Shared & 0.105*** & 0.102*** & 0.042 & 0.040 \\
 & (0.028) & (0.029) & (0.039) & (0.039) \\
\midrule
\multicolumn{5}{@{}l}{\textit{First Stage Diagnostics}} \\
\addlinespace[0.1cm]
F-statistic & 670.7 & 664.6 & 443.7 & 446.2 \\
\midrule
Observations & 9,035 & 9,035 & 9,486 & 9,486 \\
Controls & No & Yes & No & Yes \\
\bottomrule
\bottomrule
\end{tabular}
}
\caption*{\footnotesize{\textit{Note: Instrumental variables estimates of the effect of credential sharing on employment, separately for high-RATE and low-RATE groups. High-RATE learners are those with above-median predicted treatment effects on credential sharing based on a Generalized Random Forest model; low-RATE learners are those below the median. The treatment indicator serves as an instrument for credential sharing. Columns 1 and 3 present estimates without controls; columns 2 and 4 include controls for education, country, skill scores, and other pre-treatment characteristics. Baseline (\%) reports the employment rate in the control group for each RATE group. The results show that learners most responsive to the intervention (high-RATE) also benefit most from credential sharing. Standard errors in parentheses. }}}
\end{table}

\subsection{Credential Visibility in the Full Sample}
\label{sec:views}

The analyses so far used the \emph{LinkedIn Matched Sample} to examine credential sharing and employment outcomes. We now return to the full Coursera sample---which includes learners in the LinkedIn matched sample as well as others, including those without LinkedIn accounts---to examine the intervention's effect on credential visibility. While we cannot observe employment outcomes for the full sample, view counts provide a proxy for credential visibility to potential employers and allow us to validate that the intervention operated as intended across the entire experimental population.

Table \ref{clicks_1} presents OLS estimates of the treatment effect on the probability of receiving at least one view during the measurement window. We examine four outcome measures: views from any source (Column 1), views from any source excluding the learner's own views (Column 2), views originating from LinkedIn (Column 3), and LinkedIn views excluding self-views (Column 4). The distinction between all views and non-self views isolates visibility to external parties rather than learners monitoring their own profiles.

\begin{table}[!htbp] \centering 
  \caption{Impact of Credential Feature on the probability of receiving views} 
      \resizebox{0.8\textwidth}{!}{%
\begin{tabular}{@{}lcccc@{}}
\toprule
\toprule
                                 & \multicolumn{4}{c}{OLS}  \\ \cmidrule(l){2-5}
                                 & All views & All views by others & Views LinkedIn & Views LinkedIn by others \\ \cmidrule(l){2-5}
ATE                              & 0.00619 & 0.00246 & 0.00600 & 0.00214 \\
                                 & (0.00090) & (0.00080) & (0.00085) & (0.00075) \\
ATE \% baseline                  & 3.371 & 1.8041 & 3.8036 & 1.8278 \\
                                 & (0.5473) & (0.4894) & (0.5152) & (0.4596) \\
Baseline                         & 0.1884 & 0.1421 & 0.1621 & 0.1228 \\
                                 & (0.00063) & (0.00056) & (0.00059) & (0.00053) \\
                                 \hline \\[-1.8ex] 
 Observations  &  765,616 &  765,616 &  765,616 &   765,616 \\
  Learners covariates  &  Yes &  Yes &  Yes &   Yes \\
\bottomrule \bottomrule
\end{tabular}

}
  \caption*{\footnotesize{\textit{Note: Estimates of the average treatment effect on the probability of receiving views from LinkedIn. Columns 1 and 3 have all LinkedIn views as outcomes, and Columns 2 and 4 restrict attention to views not by the user. Estimates in Columns 1 and 2 are from the OLS estimator, and in Columns 3 and 4, they are with logit regression. Each estimate is adjusted using learners' characteristics as controls.
  }}}\label{clicks_1}
\end{table}

The \textit{Credential Feature} generates statistically significant increases in visibility across all measures. Treated learners are 0.62 percentage points more likely to receive at least one view from any source (Column 1), a 3.4\% increase from the baseline of 18.8\%. Excluding self-views (Column 2), the effect is 0.25 percentage points, or 1.8\% from baseline. When restricting to views originating from LinkedIn---where the platform's algorithms and professional network structure make credentials particularly salient---the effects are 0.60 percentage points (3.8\% increase) for all LinkedIn views and 0.21 percentage points (1.8\% increase) for LinkedIn views excluding self-views.

The consistency of effects across view definitions indicates that the intervention genuinely increases visibility to external viewers rather than merely prompting learners to check their own profiles more frequently. All estimates are statistically significant at conventional levels and include controls for baseline learner characteristics. However, the magnitudes of these estimates is substantially smaller than the magnitudes within the \emph{LinkedIn Matched Sample} suggesting that many users might not have accounts on professional networking platforms, do not actively use it, or where not looking for a job in the first place.

%% file: 5.resume_scoring.tex
\section{Quantifying Credential Salience: Resume Scoring Analysis}
\label{sec:resume_scoring}

Our findings demonstrate that credential sharing improves employment outcomes, with effects concentrated among learners with weak baseline signals. This pattern suggests that credentials increase profile salience to employers—making job seekers more visible and attractive during the screening process. To provide direct evidence for this mechanism and quantify its magnitude, we conduct a resume scoring exercise that simulates recruiter evaluation of LinkedIn profiles.

This analysis addresses three questions. First, do credentials substantively improve resume quality as perceived by recruiters? Second, which types of resumes benefit most from credential addition? Third, what is the magnitude of improvement a credential provides in the screening process?

\subsection{Methodology}

We employ an LLM-as-judge framework \citep{zheng2023judging} using Resume-Matcher \citep{resume_matcher}, an open-source tool that prompts a large language model to evaluate the fit between a candidate profile and a specific job opening. The system outputs a resume score from 0 to 100 indicating match quality. This approach simulates the initial screening process that job seekers face when evaluated by applicant tracking systems or recruiters.

\paragraph{Job Descriptions and Scoring Procedure.} To approximate realistic evaluation contexts, we simulate job descriptions using GPT-5 that match the primary domain of each learner's credential (Data Science, Business, Computer Science, Information Technology). Job descriptions for each domain are provided in Appendix \ref{app:job_descriptions}.

For each learner, we prepare two variants of their LinkedIn profile. The \textit{baseline resume} constructs a pre-treatment snapshot by removing: (i) the earned Coursera credential and any other certificates added after randomization, (ii) employment positions with start dates after the experiment began, and (iii) any profile elements with timestamps after the intervention date. This captures the profile as it appeared at randomization, before learners could display their newly earned credentials. The \textit{credential-enhanced resume} adds only the Coursera credential earned during the study period, isolating the marginal effect of credential visibility.

We score each learner's baseline resume twice against domain-matched job descriptions to assess scoring reliability, then score the credential-enhanced resume once. This yields three scores per learner: two baseline scores from independent runs and one score with the credential added.

\subsection{Results}

\begin{figure}[htbp]
\centering
\caption{Distribution of Resume Scores: Baseline and Credential-Enhanced Profiles}
\includegraphics[width=0.65\textwidth]{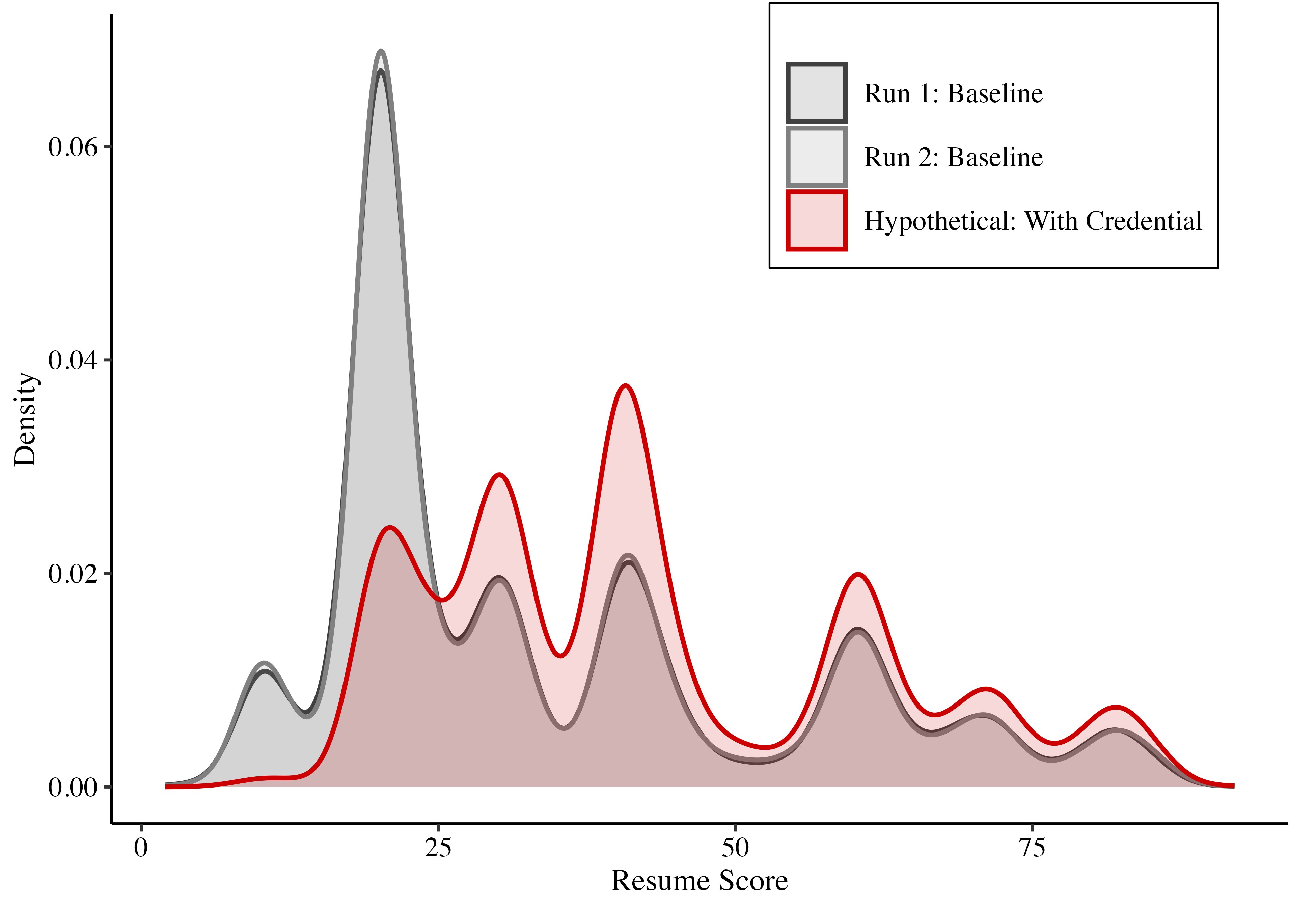}
\label{fig:score_distributions}
\caption*{\footnotesize{\textit{Notes: Run 1 and Run 2 show independent evaluations of baseline resumes without credentials. Hypothetical shows scores after adding earned credentials. All scores generated using Llama 3 via Resume-Matcher tool.}}}
\end{figure}

\paragraph{Scoring Reliability and Distributional Effects.} Figure \ref{fig:score_distributions} displays kernel density estimates of resume scores across the three scoring runs. Run 1 and Run 2 (gray) represent independent evaluations of baseline resumes, while the Hypothetical distribution (red) shows scores after adding credentials.

Run 1 and Run 2 overlap closely. A Kolmogorov-Smirnov test confirms the distributions are statistically indistinguishable ($D = 0.004$, $p = 0.998$), indicating consistent scoring of identical profiles. This reliability is essential for interpreting subsequent differences as genuine credential effects rather than measurement noise.

The Hypothetical distribution diverges markedly from the baseline runs ($D = 0.310$, $p < 0.001$). Three patterns emerge. First, baseline distributions show pronounced concentration around 20, with secondary modes near 40. Second, adding credentials shifts mass away from the lowest scores—the sharp peak at 20 is noticeably attenuated in the Hypothetical distribution, while mass increases in the 30-50 range. This rightward shift indicates credentials provide the largest absolute improvements for initially weak profiles. Third, differences between distributions are modest at higher resume scores, suggesting that for learners with strong baseline resumes, credentials do not substantially shift recruiter perceptions.

\paragraph{Validation: Resume Scores Predict Employment Outcomes.} To validate that resume scores capture meaningful employability signals, we restrict our sample to control group learners who did not share certificates ($N = 11{,}712$). Observed employment outcomes thus reflect only pre-existing profile characteristics, not credential effects.

Table \ref{tab:score_validation} presents results from OLS and logistic regression of employment outcomes on baseline resume scores. As a sanity check, we examine whether resume scores predict LinkedIn profile views—a measure of employer attention that should \textit{not} correlate with baseline scores absent credential sharing.

\begin{table}[!htbp] \centering 
  \caption{Validation: Resume Scores Predict Employment Outcomes} 
  \resizebox{0.5\textwidth}{!}{%
\begin{tabular}{@{}lcccc@{}}
\toprule
\toprule
                                 & \multicolumn{2}{c}{New Job} & \multicolumn{2}{c}{LinkedIn Views} \\ \cmidrule(lr){2-3} \cmidrule(lr){4-5}
                                 & OLS & Logit & OLS & Logit \\ \cmidrule(lr){2-3} \cmidrule(lr){4-5}
Resume Score                     & 0.006*** & 0.051*** & $-$0.00001 & $-$0.0001 \\
                                 & (0.0002) & (0.002) & (0.0002) & (0.001) \\
Constant                         & $-$0.083*** & $-$4.036*** & 0.266*** & $-$1.014*** \\
                                 & (0.005) & (0.073) & (0.008) & (0.042) \\
\midrule
Observations                     & 11,712 & 11,712 & 11,712 & 11,712 \\
\bottomrule \bottomrule
\end{tabular}
}
  \caption*{\footnotesize{\textit{Note: Sample restricted to control group learners who did not share certificates. Resume scores generated by Llama 3 via Resume-Matcher for baseline profiles without credentials. Columns 1-2 examine employment outcomes; columns 3-4 examine LinkedIn profile views as a sanity check. Standard errors in parentheses. ***p$<$0.01, **p$<$0.05, *p$<$0.1.}}}
  \label{tab:score_validation}
\end{table}

Resume scores strongly predict employment. The logistic regression coefficient of 0.051 (SE = 0.002) indicates that a one standard deviation increase in resume score (19.5 points) raises the odds of finding a new job by 12 percentage points. The mean improvement from adding credentials (8.5 points) translates to a 5.1 percentage point increase in employment odds. In contrast, resume scores show no relationship with LinkedIn profile views (columns 3-4), confirming that higher scores do not mechanically generate employer attention absent actual credential changes.

\paragraph{Heterogeneous Effects by Observable Characteristics.} To investigate which learners benefit most from credential visibility, we examine heterogeneous effects across four dimensions: baseline resume quality, prior work experience in the credential domain, predicted employability, and estimated conditional average treatment effects (CATE). For resume quality, we classify learners as having ``poor resumes'' using scores from Run 2 (33rd percentile threshold), then estimate impacts using scores from Run 1. This sample-splitting procedure ensures the same score is not used both to define subgroups and measure treatment effects.

Table \ref{tab:heterogeneous_effects} presents results. For each comparison, we report baseline employment rates and mean changes in resume scores. The final column shows the difference in mean score changes between groups.

\begin{table}[!htbp] \centering 
  \caption{Heterogeneous Effects of Credentials Across Learner Characteristics} 
  \resizebox{0.9\textwidth}{!}{%
\begin{tabular}{@{}lccccccc@{}}
\toprule
\toprule
 & \multicolumn{3}{c}{Group A (Baseline)} & \multicolumn{3}{c}{Group B (Treatment-Responsive)} & Difference \\ 
\cmidrule(lr){2-4} \cmidrule(lr){5-7} \cmidrule(lr){8-8}
Characteristic & Baseline & $\Delta$ Score & SE & Baseline & $\Delta$ Score & SE & $\Delta\Delta$ Score (SE) \\ 
\midrule
\textit{Panel A: Resume Quality} \\
Strong Resume (p67+) & 0.182 & 8.31 & (0.11) & & & &  \\
Poor Resume (p33-) & & & & 0.070 & 9.67 & (0.31) & 1.36*** \\
 & & & & & & & (0.33) \\
\midrule
\textit{Panel B: Domain Experience} \\
No Past Job in Domain & 0.123 & 8.99 & (0.11) & & & &  \\
Past Job in Domain & & & & 0.418 & 5.60 & (0.26) & $-$3.39*** \\
 & & & & & & & (0.29) \\
\midrule
\textit{Panel C: Predicted Employability} \\
Low Employability & 0.117 & 9.03 & (0.12) & & & &  \\
High Employability & & & & 0.288 & 7.15 & (0.19) & $-$1.89*** \\
 & & & & & & & (0.22) \\
\midrule
\textit{Panel D: Estimated CATE} \\
Low CATE & 0.169 & 8.17 & (0.20) & & & &  \\
High CATE & & & & 0.169 & 8.57 & (0.21) & 0.40 \\
 & & & & & & & (0.29) \\
\bottomrule \bottomrule
\end{tabular}
}
  \caption*{\footnotesize{\textit{Note: Heterogeneous effects of credential addition on resume scores across learner subgroups. Baseline reports employment rate in each group. $\Delta$ Score is the mean change in resume score when adding credentials. $\Delta\Delta$ Score is the difference in mean score changes between groups. For Panel A, poor resume classification uses Run 2 scores (cross-fitting) while impacts are measured using Run 1 scores. Standard errors in parentheses. ***p$<$0.01, **p$<$0.05, *p$<$0.1.}}}
  \label{tab:heterogeneous_effects}
\end{table}

The results reveal substantial heterogeneity. Learners with poor baseline resumes gain 1.36 more points than those with strong resumes (p $<$ 0.01). Learners without prior domain experience gain 3.39 more points than those with relevant work history (p $<$ 0.01). Low-employability learners gain 1.89 more points than high-employability learners (p $<$ 0.01). In contrast, estimated CATE shows no significant difference (0.40 points, p = 0.19), suggesting observable characteristics capture most meaningful heterogeneity in treatment effects.

\begin{figure}[!htbp]
\centering
\caption{Heterogeneous Credential Effects by Baseline Resume Quality}
\includegraphics[width=0.75\textwidth]{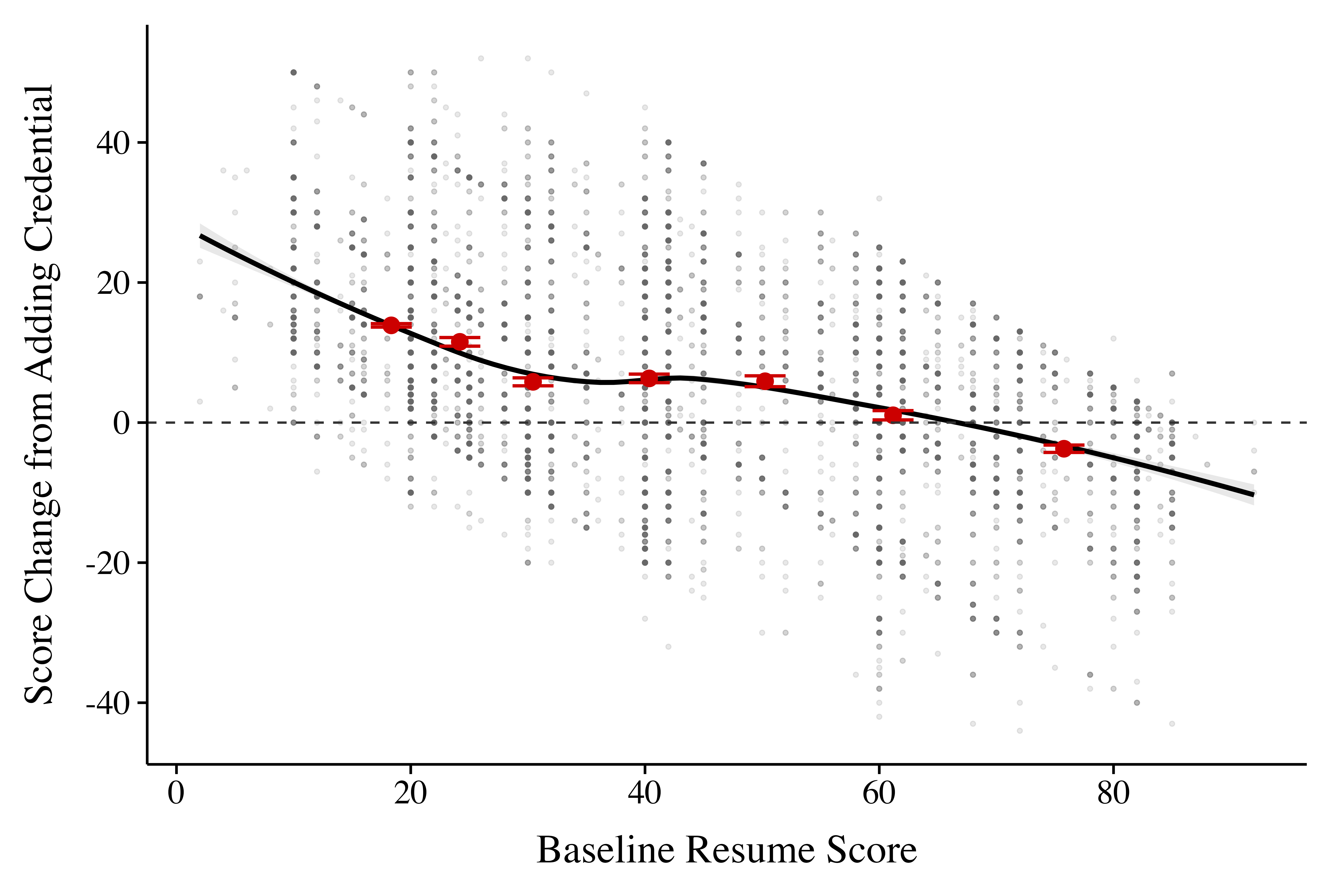}
\label{fig:score_diff_by_baseline}
\caption*{\footnotesize{\textit{Notes: Gray points show individual observations. Red points represent decile bin means with 95\% confidence intervals. Black curve shows LOESS-smoothed conditional mean. Baseline scores from Run 1; score changes computed as difference between credential-enhanced and baseline profiles. Dashed line indicates zero effect.}}}
\end{figure}

\paragraph{Credential Value Declines with Baseline Resume Quality.} Figure \ref{fig:score_diff_by_baseline} visualizes credential effects across the baseline resume quality distribution. The figure plots score changes from adding credentials against baseline scores measured in Run 1, with red points showing decile bin means and 95\% confidence intervals. The black curve traces the conditional expectation function.

The pattern reveals a striking negative gradient. Learners with the weakest baseline resumes (scores below 20) experience credential gains averaging 15 to 20 points. In contrast, learners with strong baseline resumes (scores above 70) see minimal improvements, with effects approaching zero or turning negative. The middle range exhibits moderate gains of 7 to 10 points. This inverse relationship validates our theoretical mechanism: credentials provide salience precisely where baseline signals are weakest.

The downward slope is both economically and statistically significant. Moving from the bottom decile (baseline score $\approx$ 15) to the top decile (baseline score $\approx$ 75) reduces credential value by approximately 20 points—a three- to fourfold difference. This heterogeneity helps reconcile conflicting findings in the literature: studies sampling high-quality learners \citep{deming2016value} find modest credential returns, while our intervention—reaching learners across the quality distribution—uncovers substantial benefits for those with weak signals.

%% file: 6.conclusion.tex
\section{Conclusion}

This study provides experimental evidence on the labor market value of non-traditional credentials and the effectiveness of interventions that reduce barriers to credential sharing. We conducted a large-scale randomized experiment involving over 800,000 MOOC learners from developing countries and without college degrees, examining whether simplifying credential sharing and providing targeted encouragement improves employment outcomes.

Our findings establish three main results. First, light-touch platform interventions can substantially increase credential visibility. The Credential Feature increased credential sharing by 17 percentage points and raised the probability of new employment by 6 percent overall, with a 9 percent increase for jobs directly related to learners' certificates. When interpreted as an encouragement design, the local average treatment effect of credential sharing is 10 percentage points for any new job, indicating large employment gains for learners induced to share by the intervention.

Second, credential value is highly heterogeneous and concentrated among learners with weak baseline signals. Treatment effects are largest for learners with low predicted employability, and those most responsive to credential-sharing nudges are precisely those who benefit most from sharing. This alignment between treatment responsiveness and treatment benefits has important implications for targeting: platforms can efficiently direct interventions toward high-value populations using observable characteristics without requiring complex machine learning systems.

Third, we provide direct evidence for the salience mechanism underlying these employment effects. Using an LLM-as-judge framework to simulate recruiter evaluations, we demonstrate that credentials improve resume scores by an average of 8.5 points, with effects ranging from 15-20 points for the weakest resumes to near-zero for strong resumes. This negative gradient validates our theoretical framework: credentials function as attention-grabbing signals that fill visible gaps in profiles, with the largest effects occurring precisely where baseline signals are weakest.

These findings reconcile conflicting evidence in the literature on MOOC credential value. Prior studies sampling high-quality learners find modest returns to online credentials \citep{deming2016value}, while survey evidence from diverse learner populations reports substantial career benefits \citep{zhenghao2015whos}. Our results demonstrate that both patterns reflect genuine heterogeneity rather than measurement error. Credentials provide minimal value for workers with strong alternative signals—college degrees, extensive work history, or profiles from developed countries—but generate large employment gains for those lacking traditional credentials.

The policy implications are substantial. For educational platforms, interventions to increase credential sharing should prioritize targeting. A blanket encouragement campaign yields limited gains if most recipients already have strong signals. Targeting learners with weak baseline profiles—detectable through observable characteristics such as educational attainment, work history, and country of origin—would concentrate resources where credentials provide the largest visibility boost. For workers without traditional credentials, our findings suggest that skill signaling through online learning platforms represents a viable strategy for labor market entry, with employment gains comparable to those documented for other signaling interventions \citep{pallais2014inefficient}. For policymakers concerned with labor market inequality, the concentration of benefits among disadvantaged learners indicates that facilitating non-traditional credential sharing may reduce rather than exacerbate employment disparities. 

Several limitations warrant mention. First, our employment measures rely on self-reported LinkedIn updates, which may understate true employment effects if some treated learners do not update their profile. Second, our sample selection—learners who provided LinkedIn URLs before randomization—likely represents a more career-oriented subset of MOOC completers, limiting external validity to the broader population. Third, we measure employment outcomes 8-12 months post-intervention, but longer-run effects on career trajectories and earnings remain unknown.

Future research should examine several extensions. First, tracking long-term career outcomes would reveal whether credential sharing generates persistent employment gains or merely accelerates job finding. Second, investigating the optimal design of credential-sharing interventions across different platforms, labor markets, and credential types would provide actionable guidance for scaling these interventions.

Despite these limitations, our findings demonstrate that non-traditional credentials hold substantial labor market value for workers lacking formal qualifications, and that simple platform interventions can unlock this value at minimal cost. As online learning continues to expand globally, designing platforms to facilitate effective skill signaling may represent a scalable approach to improving employment outcomes for disadvantaged workers.

%% file: appendix.tex
\newpage{}
\setcounter{page}{1}
\gdef\thepage{A\arabic{page}}
\appendix
\section*{Appendix}

\section{Mail Spin Experiment: Summary Statistics and Covariate Balance}\label{appendix_mailspin}

\paragraph{Summary Statistics}

Table~\ref{sum_stats_mailspin} reports summary statistics for the main variables used in the analysis, separately for the treatment and control groups. Each entry shows the number of observations, mean, standard deviation, minimum, and maximum. The unit of observation is an individual learner.  

The key variables capture prior online learning activity and background characteristics. \textit{Prior Course Completions} and \textit{Prior Specialization Completions} measure the number of Coursera courses and specializations completed by each learner before the intervention. \textit{Career Goal} is an indicator equal to one if the learner explicitly stated a career-related objective on the platform. \textit{No College Degree} identifies learners without a university degree, \textit{Developing Country} flags those residing in developing economies, and \textit{Tech Domain} indicates whether the course belongs to a technology-related field. \textit{First Skill Score} measures baseline proficiency in the first assessed skill dimension (measured at course completion), while \textit{Course in English} equals one if the course language is English.  

In addition to these numeric and binary covariates, the dataset contains several categorical variables that provide richer contextual information: the learner’s \textit{country of origin}, the \textit{name of the course}, the \textit{skills taught} in the course (up to three skill tags), the \textit{primary domain} and \textit{subdomain} classification of each course, as well as metadata such as \textit{course length} and \textit{difficulty level}.

\begin{table}[!h]
\caption{Summary statistics by treatment status}\label{sum_stats_mailspin}
\centering
\resizebox{\linewidth}{!}{
\fontsize{10}{12}\selectfont
\begin{tabular}[t]{lcccccccccc}
\toprule
\multicolumn{1}{c}{ } & \multicolumn{5}{c}{Treatment} & \multicolumn{5}{c}{Control} \\
\cmidrule(l{3pt}r{3pt}){2-6} \cmidrule(l{3pt}r{3pt}){7-11}
Variable & N & Mean & SD & Min & Max & N & Mean & SD & Min & Max\\
\midrule
Prior Course Completions & 427337 & 3.36 & 6.05 & 0.0 & 877.00 & 427439 & 3.35 & 6.10 & 0.00 & 848.00\\
Prior Specialization Completions & 427337 & 0.23 & 0.79 & 0.0 & 105.00 & 427439 & 0.23 & 0.83 & 0.00 & 136.00\\
Career Goal & 427337 & 0.12 & 0.32 & 0.0 & 1.00 & 427439 & 0.12 & 0.32 & 0.00 & 1.00\\
No College Degree & 427337 & 0.67 & 0.47 & 0.0 & 1.00 & 427439 & 0.68 & 0.47 & 0.00 & 1.00\\
Developing Country & 427337 & 0.50 & 0.50 & 0.0 & 1.00 & 427439 & 0.50 & 0.50 & 0.00 & 1.00\\
Tech Domain & 427337 & 0.50 & 0.50 & 0.0 & 1.00 & 427439 & 0.50 & 0.50 & 0.00 & 1.00\\
First Skill Score & 242739 & 2.90 & 1.96 & -2.6 & 13.99 & 242538 & 2.91 & 1.97 & -2.64 & 13.71\\
Course in English & 427337 & 0.96 & 0.19 & 0.0 & 1.00 & 427439 & 0.96 & 0.19 & 0.00 & 1.00\\
\bottomrule
\end{tabular}
}
\caption*{\footnotesize{\textit{Note:  Table reports number of observations, mean, standard deviation, minimum, and maximum for each variable, separately for treated and control groups (treated = 1 vs 0).}}}
\end{table}

\begin{figure}[htp]
\caption{Difference in Covariate Values Between Treatment and Control}\label{fig:cov_balance_mailspin}
\centering
\includegraphics[scale = 0.5]{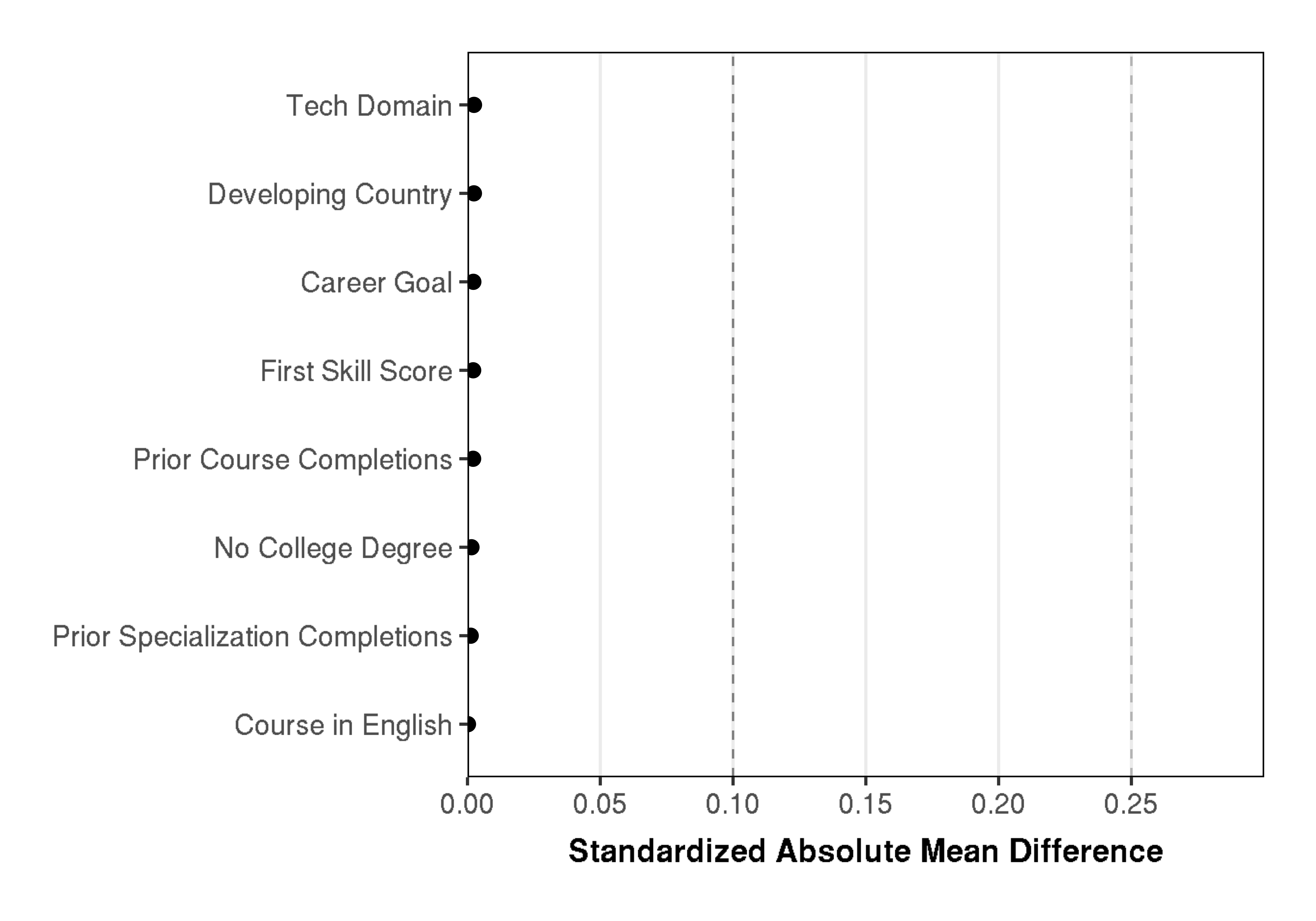}
\caption*{\footnotesize{\textit{Note: Standardized absolute mean difference between treatment and control groups in selected covariates.}}}
\end{figure}

\paragraph{Covariate Balance in Mail Spin Experiment}
Figure~\ref{fig:cov_balance_mailspin} plots standardized absolute mean differences (SMDs) for the main covariates between the treatment and control groups. Each point represents the difference in means for a given baseline variable, standardized by the pooled standard deviation. Dashed vertical lines at 0.10 and 0.25 mark thresholds for small and moderate imbalance. Across all variables, standardized differences are close to zero and well below conventional thresholds.

\section{Details of the Experimental Sample Construction}\label{ap_sample_courses}

\subsection{List of Developed Countries}\label{dev_countries}

We follow the OECD list of developed countries. Specifically, all learners with an account registered in any of the following countries are labeled as 'developed country': Austria, Australia, Belgium, Canada, Chile, Colombia, Costa Rica, Czech Republic, Denmark, Estonia, Finland, France, Germany, Greece, Hungary, Iceland, Ireland, Israel, Italy, Japan, Korea, Republic of, Latvia, Lithuania, Luxembourg, Mexico, Netherlands, New Zealand, Norway, Poland, Portugal, Slovak Republic, Slovenia, Spain, Sweden, Switzerland, Turkey, United Kingdom,  United States. 

Learners that do not come from any of these countries are consider as learners from developing countries.

\subsection{Sample of Courses Included in the Experiment}

In Table \ref{sample_courses}, we show a sample of 50 courses of 7355 from which the learners included in the experiment graduated. 

\begin{table}[ht]
\centering
  \caption{Sample of course included in the experiment} 
        \resizebox{0.9\textwidth}{!}{%
\begin{tabular}{lll}
\\[-1.8ex]\hline 
\hline \\[-1.8ex] 
  Primary domain & Credential Type & Certificate Name \\ 
  \hline
  Information Technology & Course & Fundamentos do Suporte Técnico \\ 
  Business & Guided Project & Develop a Company Website with Wix \\ 
  Data Science & Course & SQL for Data Science \\ 
  Information Technology & Course & Fundamentals of Red Hat Enterprise Linux \\ 
  Data Science & Course & Neural Networks and Deep Learning \\ 
  Information Technology & Course & Introduction to Cloud Computing \\ 
  Information Technology & Course & AWS Cloud Practitioner Essentials \\ 
  Business & Course & Capital-investissement et capital-risque \\ 
  Data Science & Specialization & Introduction to Data Science \\ 
  Business & Course & Teamwork Skills: Communicating Effectively in Groups \\ 
  Information Technology & Course & Crash Course on Python \\ 
  Business & Course & Excel Skills for Business: Advanced \\ 
  Data Science & Guided Project & Introduction to Business Analysis Using Spreadsheets: Basics \\ 
  Business & Course & Foundations of Project Management \\ 
  Business & Course & Assess for Success: Marketing Analytics and Measurement \\ 
  Business & Course & Bookkeeping Basics \\ 
  Computer Science & Course & Introduction to Front-End Development \\ 
  Business & Guided Project & Create a Project Charter with Google Docs \\ 
  Data Science & Professional Certificate & Google Data Analytics \\ 
  Information Technology & Course & Technical Support Fundamentals \\ 
  Computer Science & Course & Python Programming: A Concise Introduction \\ 
  Information Technology & Course & Introduction to Web Development with HTML, CSS, JavaScript \\ 
  Computer Science & Guided Project & Get Started with Figma \\ 
  Computer Science & Course & Foundations of User Experience (UX) Design \\ 
  Computer Science & Course & Programming for Everybody (Getting Started with Python) \\ 
  Data Science & Professional Certificate & IBM Data Analyst \\ 
  Computer Science & Course & JavaScript Basics \\ 
  Business & Course & Foundations of Digital Marketing and E-commerce \\ 
  Data Science & Course & Foundations: Data, Data, Everywhere \\ 
  Information Technology & Course & AWS Cloud Technical Essentials \\ 
  Computer Science & Course & Blockchain: Foundations and Use Cases \\ 
  Computer Science & Course & HTML, CSS, and Javascript for Web Developers \\ 
  Business & Course & Developing Innovative Ideas for Product Leaders \\ 
  Business & Guided Project & Introduction to Microsoft Excel \\ 
  Business & Course & Construction Project Management \\ 
  Data Science & Course & Introduction to Genomic Technologies \\ 
  Information Technology & Course & Explore Core Data Concepts in Microsoft Azure \\ 
  Computer Science & Course & Responsive Website Basics: Code with HTML, CSS, and JavaScript \\ 
  Business & Course & Esports Teams and Professional Players \\ 
  Computer Science & Guided Project & Build a mobile app with Google Sheets on Glide and no coding \\ 
  Business & Guided Project & Designing and Formatting a Presentation in PowerPoint \\ 
\\[-1.8ex]\hline 
\hline \\[-1.8ex] 
\end{tabular}
}
    \caption*{\footnotesize{\textit{Note: A sample of courses in which learners' included in the experiment graduated from.}}}\label{sample_courses}
\end{table}

\subsection{Geographical Distribution of Experimental Subjects}\label{map_world}

\begin{figure}[htp]
\centering
\caption{Geographical Distribution of Experimental Subjects.} 
\includegraphics[scale = 0.5]{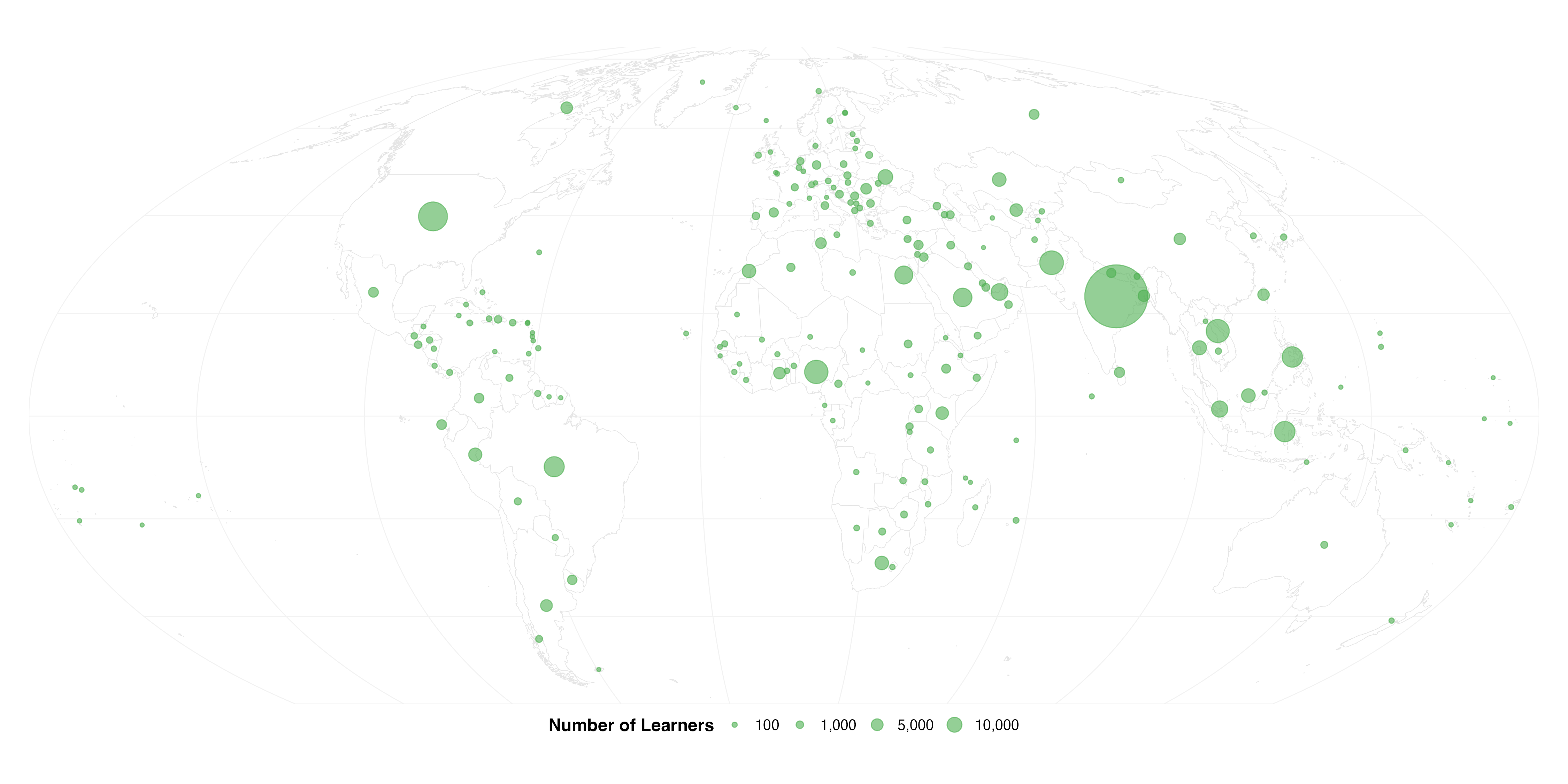}
\caption*{\footnotesize{\textit{Note:} The figure shows geographical distribution of experimental subjects. The size of circles corresponds to the number of subjects in a country.}}
\label{fig:world_map}
\end{figure}

Figure \ref{fig:world_map} displays the global distribution of learners who participated in the experiment. Each circle represents the number of participants from a given country, with the size of the circle proportional to the number of observations. The sample spans over one hundred countries, with particularly high concentrations of learners from India, the United States, Brazil, and several European countries. Participation from African, East Asian, and Latin American countries is also substantial, illustrating the broad international reach of the experiment.  

\subsection{LinkedIn Feature Engineering}

\label{app:linkedin_features}

\begin{itemize}
\item \textbf{Current Enrollment in Educational Program}: This binary feature is set to \textit{TRUE} if the end date of the participant's educational program is later than 2022. It is important to note that only the year of the start and end dates of the educational programs are available in our dataset.
\item \textbf{Level of Education}: This categorical feature classifies the participant's highest level of education based on keywords found in the title of their degree. The classifications are as follows:
\begin{itemize}
    \item \textit{Master’s Degree}: Identified through keywords such as 'master', 'msc', 'maestría', or 'ma'.
    \item \textit{Bachelor’s Degree}: Identified through keywords like 'bsc', 'bs', 'bachillerato', or 'bachelor'.
    \item \textit{Doctor’s Degree}:Identified through keywords such as 'doctor', 'doutorado', or 'docteur'.
    \item \textit{Degree}: This is marked as \textit{TRUE} if any of the above conditions are satisfied.
\end{itemize}

\item \textbf{University Ranking of the Latest Academic Degree}: Utilizing the national rankings provided by a public Kaggle dataset (\url{https://www.kaggle.com/datasets/mylesoneill/world-university-rankings}), we assigned rankings based on the latest available data, predominantly from 2015. Note that rankings may vary annually, and institutions may hold different positions in different years.

\item \textbf{Years Since Latest Academic Degree}: This feature calculates the difference in years between 2023 and the year of the participant's most recent academic degree. If the result is negative, indicating that the participant has not yet graduated, the value is set to 0.
\end{itemize}

Additionally, to discern career outcomes from the LinkedIn scraped data, the following methodology was applied:

\begin{enumerate}
\item \textbf{Internship Identification}: Initially, we extracted the current job position from the profile to ascertain whether the term “Intern” was present, which would indicate an internship role.

\item \textbf{Time Difference Calculation}: Subsequently, we calculated the time difference in months between the start date of the most recent experience and September 2022, when our experiment started.

\item \textbf{Career Outcome Determination}: Based on the computed time difference, we categorized the career outcomes as follows:
\begin{itemize}
    \item If the time difference is greater than or equal to 0 months, it implies a recent career development, leading to further analysis:
    \begin{itemize}
        \item If “Intern” is found in the current job title, the outcome is classified as a \textit{new internship}.
        \item If the employer of the current job differs from the employer of the previous job, it indicates a change in job roles, leading to the classification of \textit{new job}.
        \item Otherwise, the outcome is classified as a \textit{promotion}.
    \end{itemize}
\end{itemize}
\end{enumerate}

\section{Descriptive statistics}\label{descriptives_appendix}

Table  \ref{tab:covariate_coursera_table} presents summary statistics for the covariates provided by the Coursera internal dataset.

\begin{table}[H]
\caption{Summary Statistics for Internal Coursera Covariates} 
\centering
\resizebox{0.7\textwidth}{!}{%
\begin{tabular}{llrrrrr}
\\[-1.8ex]\hline 
\hline \\[-1.8ex] 
Variable & Group & Count & Min & Max & Mean & SD \\ 
  \hline
gender & unknown & 470459 & 0.00 & 1.00 & 0.53 & 0.50 \\ 
  gender & male & 269836 & 0.00 & 1.00 & 0.30 & 0.46 \\ 
  gender & female & 146897 & 0.00 & 1.00 & 0.17 & 0.37 \\ 
  gender & other & 1027 & 0.00 & 1.00 & 0.00 & 0.03 \\ 
  education level & unknown & 570268 & 0.00 & 1.00 & 0.64 & 0.48 \\ 
  education level & associate degree & 25298 & 0.00 & 1.00 & 0.03 & 0.17 \\ 
  education level & masters degree & 44975 & 0.00 & 1.00 & 0.05 & 0.22 \\ 
  education level & bachelor degree & 114926 & 0.00 & 1.00 & 0.13 & 0.34 \\ 
  education level & professional degree & 3540 & 0.00 & 1.00 & 0.00 & 0.06 \\ 
  education level & college no degree & 64078 & 0.00 & 1.00 & 0.07 & 0.26 \\ 
  education level & high school diploma & 53155 & 0.00 & 1.00 & 0.06 & 0.24 \\ 
  education level & doctorate degree & 3963 & 0.00 & 1.00 & 0.00 & 0.07 \\ 
  education level & less than high school diploma & 8016 & 0.00 & 1.00 & 0.01 & 0.09 \\ 
  primary domain & business & 330462 & 0.00 & 1.00 & 0.37 & 0.48 \\ 
  primary domain & data science & 208819 & 0.00 & 1.00 & 0.24 & 0.42 \\ 
  primary domain & computer science & 222335 & 0.00 & 1.00 & 0.25 & 0.43 \\ 
  primary domain & information technology & 126603 & 0.00 & 1.00 & 0.14 & 0.35 \\ 
  credential type & course & 727602 & 0.00 & 1.00 & 0.82 & 0.38 \\ 
  credential type & guided project & 147540 & 0.00 & 1.00 & 0.17 & 0.37 \\ 
  credential type & specialization & 8278 & 0.00 & 1.00 & 0.01 & 0.10 \\ 
  credential type & professional certificate & 4799 & 0.00 & 1.00 & 0.01 & 0.07 \\ 
  developed country & - & 888219 & 0.00 & 1.00 & 0.10 & 0.30 \\ 
  certificate has page views & - & 888219 & 0.00 & 1.00 & 0.20 & 0.40 \\ 
  certificate has page views from linkedin & - & 888219 & 0.00 & 1.00 & 0.17 & 0.38 \\ 
  count all views & - & 888219 & 0.00 & 726.00 & 0.71 & 3.09 \\ 
  count all views not by user & - & 888219 & 0.00 & 725.00 & 0.60 & 3.03 \\ 
  count linkedin views & - & 888219 & 0.00 & 411.00 & 0.58 & 2.37 \\ 
  count linkedin views not by user & - & 888219 & 0.00 & 411.00 & 0.49 & 2.31 \\ 
  has degree linkedin & - & 20396 & 0.00 & 1.00 & 0.62 & 0.48 \\ 
  has bachelor linkedin & - & 20396 & 0.00 & 1.00 & 0.43 & 0.49 \\ 
  has master linkedin & - & 20396 & 0.00 & 1.00 & 0.18 & 0.38 \\ 
  has doctor linkedin & - & 20396 & 0.00 & 1.00 & 0.02 & 0.13 \\ 
  yearsSinceEnrollment & - & 20396 & 0.00 & 46.00 & 3.97 & 5.43 \\ 
  new internship linkedin & - & 20396 & 0.00 & 1.00 & 0.05 & 0.22 \\ 
  new job linkedin & - & 20396 & 0.00 & 1.00 & 0.31 & 0.46 \\ 
  promotion linkedin & - & 20396 & 0.00 & 1.00 & 0.08 & 0.27 \\ 
  new job or promotion linkedin & - & 20396 & 0.00 & 1.00 & 0.39 & 0.49 \\ 
  any outcome linkedin & - & 20396 & 0.00 & 1.00 & 0.44 & 0.50 \\ 
\\[-1.8ex]\hline 
\hline \\[-1.8ex] 
\end{tabular}
}
\label{tab:covariate_coursera_table}
\end{table}

\subsection{Balance check analysis}\label{balance_appendix}

Table \ref{tab:cov_balance} presents a balance check analysis for the total population of the experiment participants. The first column shows the difference between means in treatment and control; second column the standard error of the difference, and columns 3 and 4 means in treatment and control. We find that all the two groups are balanced in all covariates.

\begin{table}[H]
\caption{Covariate balance between treatment and control}\label{tab:cov_balance}
\centering
\renewcommand{\arraystretch}{0.8} 
\resizebox{0.999\textwidth}{!}{%
\begin{tabular}{@{\extracolsep{5pt}}lccccccc} 
\toprule
\toprule
Variable & Mean Difference & Standard Error & Treatment Mean & Control Mean & Treatment N & Control N \\ 
\hline \\[-1.8ex] 
First Skill Score  & 0.0003 & 0.0014 & 0.1002 & 0.0999 & 381{,}815 & 383{,}801 \\ 
First Skill Score Standardized & 0.0000 & 0.0001 & 0.0000 & 0.0000 & 381{,}815 & 383{,}801 \\ 
Associate Degree & 0.0003 & 0.0004 & 0.0290 & 0.0287 & 381{,}815 & 383{,}801 \\ 
Bachelor’s Degree & $-0.0008$ & 0.0008 & 0.1269 & 0.1278 & 381{,}815 & 383{,}801 \\ 
Some College & $-0.0005$ & 0.0006 & 0.0721 & 0.0726 & 381{,}815 & 383{,}801 \\ 
Doctorate Degree & $-0.0002$ & 0.0002 & 0.0043 & 0.0045 & 381{,}815 & 383{,}801 \\ 
High School Diploma & $-0.0004$ & 0.0005 & 0.0592 & 0.0596 & 381{,}815 & 383{,}801 \\ 
Less Than High School & 0.0000 & 0.0002 & 0.0089 & 0.0090 & 381{,}815 & 383{,}801 \\ 
Master’s Degree & $-0.0007$ & 0.0005 & 0.0496 & 0.0503 & 381{,}815 & 383{,}801 \\ 
No Education Mentioned & 0.0022 & 0.0011 & 0.6458 & 0.6436 & 381{,}815 & 383{,}801 \\ 
Professional Degree & 0.0002 & 0.0001 & 0.0041 & 0.0038 & 381{,}815 & 383{,}801 \\ 
Male & $-0.0008$ & 0.0010 & 0.3014 & 0.3022 & 381{,}815 & 383{,}801 \\ 
Gender Not Mentioned & 0.0009 & 0.0011 & 0.5336 & 0.5328 & 381{,}815 & 383{,}801 \\ 
Primary Domain: Computer Science & 0.0015 & 0.0010 & 0.2529 & 0.2514 & 381{,}815 & 383{,}801 \\ 
Primary Domain: Data Science & $-0.0010$ & 0.0010 & 0.2353 & 0.2363 & 381{,}815 & 383{,}801 \\ 
Primary Domain: Information Technology & $-0.0006$ & 0.0008 & 0.1398 & 0.1404 & 381{,}815 & 383{,}801 \\ 
Credential Type: Guided Project & 0.0001 & 0.0009 & 0.1679 & 0.1678 & 381{,}815 & 383{,}801 \\ 
Credential Type: Professional Certificate & $-0.0001$ & 0.0002 & 0.0054 & 0.0055 & 381{,}815 & 383{,}801 \\ 
Credential Type: Specialization & $-0.0003$ & 0.0002 & 0.0084 & 0.0087 & 381{,}815 & 383{,}801 \\ 
Developing Country & 0.0005 & 0.0007 & 0.8962 & 0.8957 & 381{,}815 & 383{,}801 \\ 
Professional Experience (Years) & 0.0464 & 0.0609 & 3.6122 & 3.5658 & 18{,}487 & 18{,}459 \\ 
Past Business Job & $-0.0014$ & 0.0026 & 0.0641 & 0.0655 & 18{,}487 & 18{,}459 \\ 
Past Tech Job & 0.0034 & 0.0035 & 0.1308 & 0.1274 & 18{,}487 & 18{,}459 \\ 
\bottomrule
\bottomrule
\end{tabular} 
}
\caption*{\footnotesize{\textit{Note: Averages of covariate values across treatment and control groups.}}}
\end{table}

\subsection{Profile Completeness Analysis}
\label{appendix_profile}

This section provides additional detail on the analysis of LinkedIn profile completeness discussed in Section \ref{sec:intent_to_treat}. To test whether the \textit{Credential Feature} prompted learners to generally improve their LinkedIn profiles beyond adding credentials, we measure profile completeness using the total character count across all profile fields, excluding fields related to credentials and newly reported jobs.

\begin{figure}[!htbp]
\centering
\caption{Distribution of LinkedIn Profile Lengths by Treatment Status}\label{density_profiles}
\includegraphics[width=0.75\textwidth]{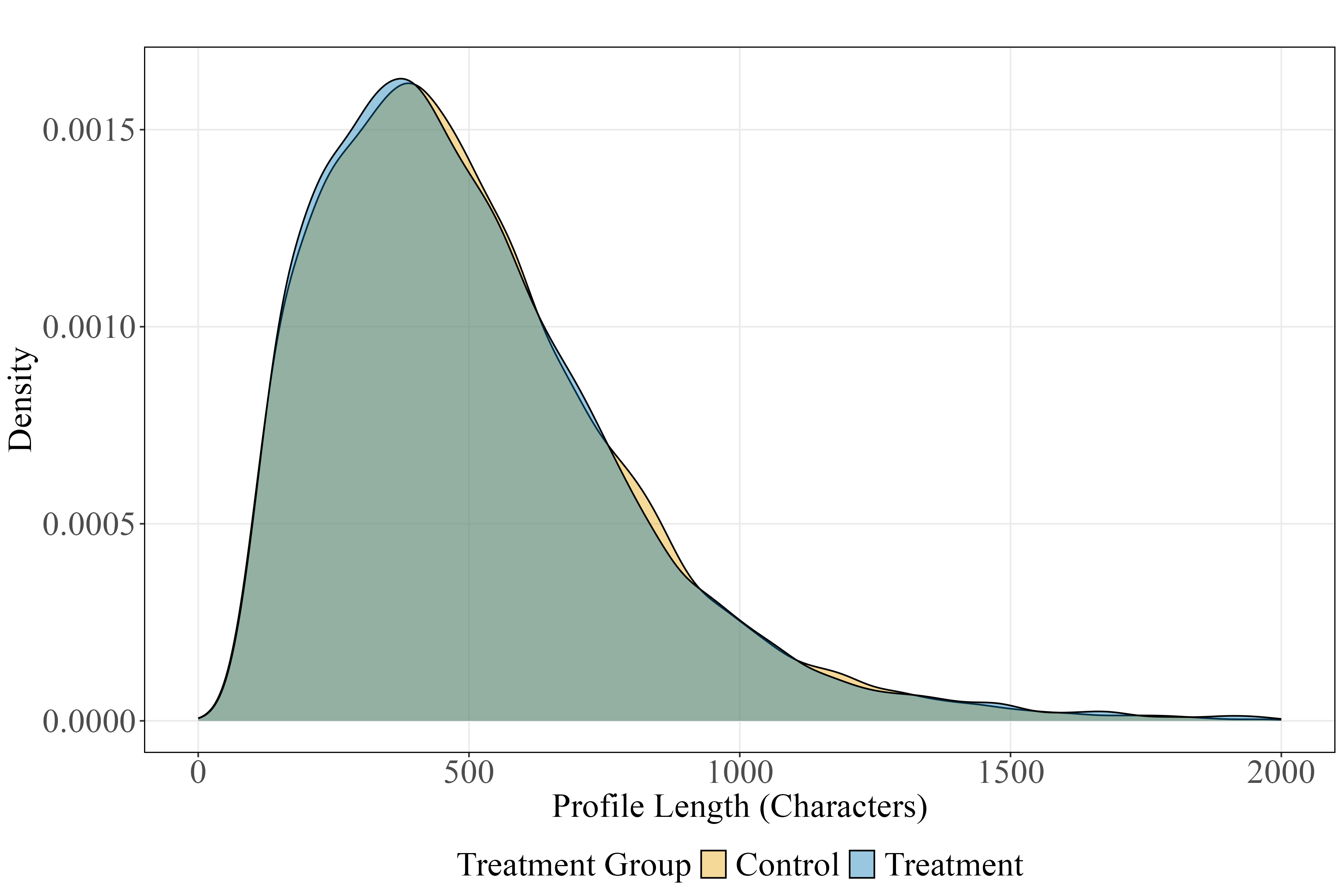}
\caption*{\footnotesize{\textit{Note:  This figure displays kernel density estimates of total profile length (measured in characters) for treatment (blue) and control (orange) groups in the LinkedIn matched sample.}}}
\end{figure}

Figure \ref{density_profiles} displays density estimates of profile length for treatment and control groups. The distributions exhibit near-perfect overlap, with both groups having similar means (512 characters for treatment, 511 for control) and nearly identical shapes. Statistical tests confirm this visual impression: a two-sample $t$-test yields $t = -0.23$ ($p = 0.82$), and a Kolmogorov-Smirnov test produces $D = 0.009$ ($p = 0.80$). Both tests fail to reject the null hypothesis of equal distributions.

\section{Details of Batched Recruitment and Results per Batch}\label{appendix_batches}

\subsection{Impact of Credential Feature on Credential Sharing}

Panel A of Table \ref{sharing_batch} presents the average treatment effect estimates on \emph{credential sharing} per batch. We can observe differences in the baseline shares of learners who shared their credentials, as well as in the average treatment effect estimates. Notably, the final batch exhibited a significantly lower treatment effect compared to other batches. Various factors may contribute to these discrepancies, including the baseline propensity to respond to treatment and the number of learners effectively exposed to the treatment.

Notifications aimed at encouraging credential sharing were displayed within the Coursera apps, hence only those learners who logged into the platform post-randomization were targeted. The percentage of learners who did so varied, with the second batch having the highest share (96\%) and the final batch having the lowest share (82\%). Furthermore, in instances where Coursera was sending additional notifications to learners, depending on their priority level, they might be shown before our notifications. In such cases, treated learners would encounter the notification during their subsequent visit to the Coursera app. Panel B of Table \ref{sharing_batch} illustrates the average treatment effect on credential sharing limited to learners who logged into the app post-randomization (the restriction is applied to both treatment and control groups). We observe a higher treatment effect across all batches, including batch 5, compared to the unrestricted sample.

\begin{table}[!htbp] \centering 
  \caption{Average treatment effect of Credential Feature on Credential Shared by Batch} 
  \label{sharing_batch} 
  \resizebox{0.7\textwidth}{!}{%
\begin{tabular}{@{\extracolsep{5pt}}lcccccccc}
\\[-1.8ex]\hline 
\hline \\[-1.8ex]
Batch & Mean Control &   Mean Treatment &   ATE &   ATE (\%) &  \\
\hline \\[-1.8ex]
\multicolumn{6}{@{}l}{\textit{Panel A: LinkedIn Matched Sample}}\\
\addlinespace
1 & 17.832 &   20.074 &   2.242 &  12.571 &  \\
  & (0.617) &   (0.650)  & (0.896) &   (5.025) &  \\
2 & 14.992 &   18.255 &   3.262 &   21.759 &  \\
  & (0.569) &   (0.601) &   (0.828) & (5.521) &  \\
3 & 16.398 &   19.382 &   2.984 &   18.196 &  \\
  & (0.579) &   (0.627) &   (0.853) &   (5.203) &  \\
4 & 17.533 &   21.794 &   4.261 &   24.303 &  \\
  & (0.616) &   (0.669) &   (0.909) &   (5.185) &  \\
5 & 16.546 &   18.123 &   1.577   & 9.528  &  \\
  & (0.706) &   (0.732) &   (1.017) & (6.147) &  \\
\midrule
\multicolumn{6}{@{}l}{\textit{Panel B: Sample of learners that logged in after randomization}}\\
\addlinespace
1 & 18.007   & 20.203   & 2.195 &   12.191 &  \\
  & (0.623)   & (0.655)  & (0.904) &   (5.022) &  \\
2 & 15.259 &   18.679 &   3.420 &   22.410 &  \\
  & (0.579) &   (0.614) &   (0.844) &   (5.531) &  \\
3 & 16.853 &   19.917 &   3.064 &   18.183 &  \\
  & (0.596) &   (0.644) &   (0.877) &   (5.206) &  \\
4 & 17.749 &   22.292 &   4.543 &   25.594 &  \\
  & (0.628) &   (0.686) & (0.930) &   (5.241) &  \\
5 & 16.829 &   18.581 &   1.752 &   10.411 &  \\
  & (0.724) &   (0.754) &   (1.045) &   (6.211) &  \\
\\[-1.8ex]\hline 
\hline \\[-1.8ex] 
\end{tabular}

}
\caption*{\footnotesize{\textit{Note: Estimates of the average treatment effect obtained using a difference-in-means estimator. Standard errors in parentheses.}}} 
\end{table}

\section{Correlation between views and new jobs}\label{cor_views_jobs}

Table \ref{corr_views_jobs} shows the estimates from the linear probability regressions of \emph{New job} on the four types of credential views using LinkedIn Matched Sample. We find that there is a strong correlation between receiving credential views and reporting a new job.

\begin{table}[!htbp] \centering 
  \caption{Correlation between views and new jobs} 
  \label{corr_views_jobs} 
  \resizebox{0.99\textwidth}{!}{%
\begin{tabular}{@{\extracolsep{5pt}}lcccc} 
\\[-1.8ex]\hline 
\hline \\[-1.8ex] 
 & \multicolumn{4}{c}{\textit{Dependent variable:}} \\ 
\cline{2-5} 
\\[-1.8ex] & All views & All views by others & Views LinkedIn & Views LinkedIn by others \\ 
\\[-1.8ex] & (1) & (2) & (3) & (4)\\ 
\hline \\[-1.8ex] 
 New job & 0.513$^{***}$ & 0.400$^{***}$ & 0.497$^{***}$ & 0.379$^{***}$ \\ 
  & (0.008) & (0.007) & (0.007) & (0.006) \\ 
  & & & & \\ 
\hline \\[-1.8ex] 
Observations & 36,946 & 36,946 & 36,946 & 36,946 \\ 
R$^{2}$ & 0.109 & 0.090 & 0.107 & 0.086 \\ 
Adjusted R$^{2}$ & 0.109 & 0.090 & 0.107 & 0.086 \\ 
Residual Std. Error (df = 36945) & 0.618 & 0.538 & 0.604 & 0.520 \\ 
F Statistic (df = 1; 36945) & 4,538.226$^{***}$ & 3,641.660$^{***}$ & 4,446.666$^{***}$ & 3,492.890$^{***}$ \\ 
\hline 
\hline \\[-1.8ex] 
\textit{Note:}  & \multicolumn{4}{r}{$^{*}$p$<$0.1; $^{**}$p$<$0.05; $^{***}$p$<$0.01} \\ 
\end{tabular} 
}
\caption*{\footnotesize{\textit{Note: Estimate from linear probability models regressing new jobs on the four types of views outcomes. Estimates based on the LinkedIn Matched Sample.}}}
\end{table}

\section{Sample Job Descriptions}
\label{app:job_descriptions}

To evaluate how credentials affect perceived candidate quality, we generated realistic job descriptions matching the primary domains of learners' earned certificates. This appendix presents representative examples of the job postings used in our resume scoring exercise.

\paragraph{Job Description Generation Process}

We used GPT-5 to create domain-specific job descriptions for entry- to mid-level positions across four primary credential domains: Business, Computer Science, Data Science, and Information Technology. For each domain, we instructed the model to generate realistic job postings with:
\begin{enumerate}
    \item A clear position title appropriate for early-career professionals
    \item An overview describing the role and its purpose
    \item Required qualifications reflecting typical baseline expectations
    \item Preferred qualifications indicating desirable additional skills
    \item A comprehensive list of relevant keywords and technical skills
\end{enumerate}

The job descriptions serve two purposes in our analysis. First, they provide standardized evaluation contexts for the Resume-Matcher tool, ensuring that profiles are scored against realistic employer requirements. Second, they allow us to measure how credentials affect perceived fit for positions directly aligned with the learner's area of study. Each of the 1,000 profiles in our scoring sample was evaluated against five job descriptions from their respective domain, yielding 5,000 baseline scores and 5,000 credential-enhanced scores.

\subsection{Representative Job Descriptions by Domain}

The following pages present one representative job description from each of the four primary domains. These examples illustrate the structure, level of detail, and skill requirements used across all generated postings.

\begin{mdframed}[
    linewidth=0.5pt,
    linecolor=black,
    backgroundcolor=gray!5,
    innertopmargin=10pt,
    innerbottommargin=10pt,
    innerleftmargin=10pt,
    innerrightmargin=10pt
]
{\sffamily\small
\textbf{Position: Junior Business Analyst}

\textbf{Overview:}\\
We are looking for an analytical and detail-oriented Junior Business Analyst to support our business operations. You will assist in analyzing business processes, gathering requirements, and helping drive data-informed decision making.

\textbf{Required Qualifications:}
\begin{itemize}[noitemsep,topsep=2pt]
    \item Bachelor's degree in Business Administration, Economics, Finance, or related field
    \item Strong analytical and critical thinking skills
    \item Proficiency in Microsoft Office Suite (Excel, PowerPoint, Word)
    \item Excellent written and verbal communication skills
    \item Ability to work with cross-functional teams
    \item Basic understanding of business processes and operations
    \item Strong attention to detail
\end{itemize}

\textbf{Preferred Qualifications:}
\begin{itemize}[noitemsep,topsep=2pt]
    \item Internship experience in business analysis, consulting, or related field
    \item Familiarity with data visualization tools (Tableau, Power BI)
    \item Basic SQL knowledge for data querying
    \item Experience with process mapping and documentation
    \item Understanding of project management principles
    \item Exposure to CRM or ERP systems
    \item Knowledge of business intelligence concepts
\end{itemize}

\textbf{Key Skills:}\\
Excel, PowerPoint, Data Analysis, Business Analysis, Requirements Gathering, Process Mapping, Communication, Presentation Skills, Problem Solving, SQL, Tableau, Power BI, Stakeholder Management, Documentation, Critical Thinking, Project Management, Teamwork
}
\end{mdframed}

\vspace{1cm}

\begin{mdframed}[
    linewidth=0.5pt,
    linecolor=black,
    backgroundcolor=gray!5,
    innertopmargin=10pt,
    innerbottommargin=10pt,
    innerleftmargin=10pt,
    innerrightmargin=10pt
]
{\sffamily\small
\textbf{Position: Junior Software Engineer}

\textbf{Overview:}\\
We are seeking a motivated Junior Software Engineer to join our development team. You will contribute to the design, development, and maintenance of software applications while learning from experienced engineers.

\textbf{Required Qualifications:}
\begin{itemize}[noitemsep,topsep=2pt]
    \item Bachelor's degree in Computer Science, Software Engineering, or related field
    \item Strong foundation in programming fundamentals and data structures
    \item Knowledge of at least one programming language (Python, Java, C++, or JavaScript)
    \item Understanding of algorithms and computational complexity
    \item Familiarity with version control systems (Git)
    \item Problem-solving skills and logical thinking
    \item Ability to work collaboratively in a team environment
\end{itemize}

\textbf{Preferred Qualifications:}
\begin{itemize}[noitemsep,topsep=2pt]
    \item Internship or project experience in software development
    \item Familiarity with object-oriented programming principles
    \item Understanding of software development lifecycle
    \item Experience with debugging and testing
    \item Exposure to Agile/Scrum methodologies
    \item Contributions to open source projects or personal coding projects
    \item Knowledge of web technologies (HTML, CSS, JavaScript)
\end{itemize}

\textbf{Key Skills:}\\
Python, Java, C++, JavaScript, Git, Data Structures, Algorithms, Object-Oriented Programming, Debugging, Testing, Problem Solving, Teamwork, Communication, Agile, Software Development, Version Control, Code Review
}
\end{mdframed}

\newpage

\begin{mdframed}[
    linewidth=0.5pt,
    linecolor=black,
    backgroundcolor=gray!5,
    innertopmargin=10pt,
    innerbottommargin=10pt,
    innerleftmargin=10pt,
    innerrightmargin=10pt
]
{\sffamily\small
\textbf{Position: Junior Data Scientist}

\textbf{Overview:}\\
We are seeking a Junior Data Scientist to join our analytics team. You will work on data-driven projects, perform statistical analysis, build predictive models, and help transform data into actionable insights.

\textbf{Required Qualifications:}
\begin{itemize}[noitemsep,topsep=2pt]
    \item Bachelor's or Master's degree in Data Science, Statistics, Mathematics, Computer Science, or related field
    \item Strong foundation in statistics and probability
    \item Programming skills in Python or R
    \item Knowledge of data manipulation and analysis libraries (pandas, NumPy)
    \item Understanding of machine learning fundamentals
    \item Ability to work with large datasets
    \item Strong analytical and problem-solving skills
\end{itemize}

\textbf{Preferred Qualifications:}
\begin{itemize}[noitemsep,topsep=2pt]
    \item Experience with machine learning libraries (scikit-learn, TensorFlow, PyTorch)
    \item Familiarity with data visualization tools (Matplotlib, Seaborn, Plotly)
    \item Knowledge of SQL and database querying
    \item Experience with Jupyter notebooks
    \item Understanding of A/B testing and experimental design
    \item Exposure to big data technologies (Spark, Hadoop)
    \item Portfolio of data science projects or Kaggle competitions
\end{itemize}

\textbf{Key Skills:}\\
Python, R, Statistics, Machine Learning, Data Analysis, pandas, NumPy, scikit-learn, SQL, Data Visualization, Matplotlib, Statistical Modeling, Predictive Analytics, Data Cleaning, Jupyter, Problem Solving, Communication, TensorFlow, A/B Testing, Experimentation
}
\end{mdframed}

\vspace{1cm}

\begin{mdframed}[
    linewidth=0.5pt,
    linecolor=black,
    backgroundcolor=gray!5,
    innertopmargin=10pt,
    innerbottommargin=10pt,
    innerleftmargin=10pt,
    innerrightmargin=10pt
]
{\sffamily\small
\textbf{Position: Junior IT Support Specialist}

\textbf{Overview:}\\
We are looking for a customer-focused Junior IT Support Specialist to provide technical assistance to our organization. You will troubleshoot hardware and software issues, support end-users, and help maintain our IT infrastructure.

\textbf{Required Qualifications:}
\begin{itemize}[noitemsep,topsep=2pt]
    \item Bachelor's degree in Information Technology, Computer Science, or related field (or equivalent experience)
    \item Strong understanding of computer hardware, software, and networking fundamentals
    \item Familiarity with Windows and/or macOS operating systems
    \item Basic knowledge of networking concepts (TCP/IP, DNS, DHCP)
    \item Excellent customer service and communication skills
    \item Ability to troubleshoot technical issues methodically
    \item Strong attention to detail and documentation skills
\end{itemize}

\textbf{Preferred Qualifications:}
\begin{itemize}[noitemsep,topsep=2pt]
    \item CompTIA A+, Network+, or similar IT certifications
    \item Experience with help desk ticketing systems
    \item Knowledge of Active Directory and user management
    \item Familiarity with cloud platforms (Microsoft 365, Google Workspace)
    \item Understanding of cybersecurity best practices
    \item Experience with remote desktop support tools
    \item Scripting knowledge (PowerShell, Bash)
\end{itemize}

\textbf{Key Skills:}\\
Windows, macOS, Linux, Networking, TCP/IP, Hardware Troubleshooting, Software Installation, Help Desk, Technical Support, Customer Service, Active Directory, Microsoft 365, Google Workspace, Ticketing Systems, Documentation, Problem Solving, Communication, Cybersecurity, Remote Support, CompTIA A+
}
\end{mdframed}